\documentclass[prd,nopacs,preprintnumbers,twocolumn,amsmath,nofootinbib,amssymb,floatfix,superscriptaddress]{revtex4}
\usepackage{graphicx,color,dcolumn,booktabs,bm}
\usepackage{longtable,lscape}
\usepackage{txfonts}
\usepackage{overpic}
\usepackage{amssymb}
\usepackage{indentfirst}
\usepackage{feynmf}
\usepackage{slashed}
\usepackage{cases}
\usepackage{float}
\usepackage{multirow}
\usepackage{appendix}
\usepackage[figuresright]{rotating}
\usepackage[colorlinks,
citecolor=blue,
anchorcolor=red,
menucolor=red,
linkcolor=red,
filecolor=red,
runcolor=red,
urlcolor=blue,
frenchlinks=red]{hyperref}
\usepackage{epstopdf}
\usepackage{bm}
\usepackage{tabularx}
\usepackage{threeparttable}
\usepackage{bbding}
\usepackage{array}
\usepackage[section]{placeins}
\usepackage{makecell}
\usepackage{comment}
\usepackage{mathrsfs}
\usepackage{ulem}
\usepackage{xcolor}
\usepackage{ulem}
\makeatletter

\newcommand{\Rmnum}[1]{\expandafter\@slowromancap\romannumeral #1@}
\makeatother

\begin{document}
	\title{Chiral extrapolation of the doubly charmed baryons magnetic properties}
	
	\author{Jiong-Jiong Liu}
	\affiliation{School of Physical Science and Technology, Lanzhou University, Lanzhou 730000, China}
	\affiliation{Research Center for Hadron and CSR Physics, Lanzhou University and Institute of Modern Physics of CAS, Lanzhou 730000, China}
	\affiliation{Lanzhou Center for Theoretical Physics, MoE Frontiers Science Center for Rare Isotopes, Key Laboratory of Quantum Theory and Applications of MoE, Key Laboratory of Theoretical Physics of Gansu Province, Gansu Provincial Research Center for Basic Disciplines of Quantum Physics, Lanzhou University, Lanzhou 730000, China}
	
	\author{Zhan-Wei Liu}\email{liuzhanwei@lzu.edu.cn}
	\affiliation{School of Physical Science and Technology, Lanzhou University, Lanzhou 730000, China}
	\affiliation{Research Center for Hadron and CSR Physics, Lanzhou University and Institute of Modern Physics of CAS, Lanzhou 730000, China}
	\affiliation{Lanzhou Center for Theoretical Physics, MoE Frontiers Science Center for Rare Isotopes, Key Laboratory of Quantum Theory and Applications of MoE, Key Laboratory of Theoretical Physics of Gansu Province, Gansu Provincial Research Center for Basic Disciplines of Quantum Physics, Lanzhou University, Lanzhou 730000, China}
	
	\author{Xiu-Lei Ren}\email{xiulei.ren@uni-mainz.de}
	\affiliation{ School of Nuclear Science, Energy and Power Engineering,
		Shandong University, Jinan 250061, China}
	\affiliation{Helmholtz Institut Mainz, D-55099 Mainz, Germany}
	
	\author{Yu Zhuge}
	\affiliation{School of Physical Science and Technology, Lanzhou University, Lanzhou 730000, China}
	\affiliation{Research Center for Hadron and CSR Physics, Lanzhou University and Institute of Modern Physics of CAS, Lanzhou 730000, China}
	\affiliation{Lanzhou Center for Theoretical Physics, MoE Frontiers Science Center for Rare Isotopes, Key Laboratory of Quantum Theory and Applications of MoE, Key Laboratory of Theoretical Physics of Gansu Province, Gansu Provincial Research Center for Basic Disciplines of Quantum Physics, Lanzhou University, Lanzhou 730000, China}
	
	\begin{abstract}
		The magnetic moments, magnetic form factors, and transition magnetic form factors of doubly charmed baryons are studied within heavy baryon chiral perturbation theory. We regulate the loop integrals using the finite-range regularization. The contributions of vector mesons are taken into account to investigate the dependence of form factors on the transferred momentum. The finite volume and lattice spacing effects are considered to analyze the lattice QCD simulations which can be understood well in our framework. 
	\end{abstract}

	\affiliation{}
	
	\maketitle
	
	\section{Introduction}\label{introduction}
	The physics of heavy flavor hadrons has been developing with great experimental progress and exhibited special characters due to the large masses of heavy quarks. The doubly charmed baryons have been searched for by various experimental collaborations~\cite{LHCb:2018pcs,LHCb:2019epo,SELEX:2004lln,LHCb:2021eaf,LHCb:2018zpl,LHCb:2019qed,LHCb:2019ybf,LHCb:2022rpd,Ratti:2003ez,Belle:2006edu,BaBar:2006bab,LHCb:2021rkb}. The SELEX Collaboration reported results for $\Xi_{cc}^+$ in 2002~\cite{SELEX:2002wqn}, and the LHCb Collaboration reported evidence for $\Xi_{cc}^{++}$ in 2017~\cite{LHCb:2017iph}. 
	
	The structure of these heavy flavor hadrons is important for us to understand the nonperturbative behavior of QCD. In addition to the mass spectra and strong decays, the electromagnetic form factors are indispensable tools for exploring the property of doubly charmed baryons~\cite{Wang:2018lhz,Yu:2022lel,ShekariTousi:2024mso,Aliev:2012ru,Aliev:2012iv,Wang:2010it,Wang:2022aga,Yao:2018ifh}. The magnetic moments of doubly charmed baryons were studied by utilizing the quark model since the 1970s~\cite{Lichtenberg:1976fi}, and it has been further investigated with different quark models~\cite{Barik:1983ics,Jena:1986xs,Silvestre-Brac:1996myf,Julia-Diaz:2004yqv,Albertus:2006ya,Faessler:2006ft,Majethiya:2008ia,Patel:2008xs,Dahiya:2009ix,Sharma:2010vv,Shah:2016vmd,Shah:2017liu,Rahmani:2020pol,Shah:2021reh,Mutuk:2021epz,Shah:2023mzg,Patel:2024lkj,Lai:2024jfe}, MIT bag model~\cite{Bose:1980vy,Bernotas:2012nz,Simonis:2018rld,Zhang:2021yul}, Skyrmion model~\cite{Oh:1991ws}, light-cone QCD sum rule~\cite{Ozdem:2018uue}, and so on~\cite{Kumar:2005ei,Dhir:2009ax,Hazra:2021lpa,Mohan:2022sxm,Gadaria:2016omw}. 
	
	The transition magnetic form factors connect both ground and excited doubly charmed baryons and play important roles in the radiative decays and other properties. The radiative decays of doubly charmed baryons have been studied in the MIT bag model~\cite{Hackman:1977am,Bernotas:2013eia,Simonis:2018rld}, the quark model~\cite{Sharma:2010vv}, light-cone QCD sum rule~\cite{Cui:2017udv, Aliev:2021hqq,Aliyev:2022rrf,Aliev:2023pwd}, and so on~\cite{Dhir:2009ax,Soni:2017yvw,Gadaria:2018tvt,Hazra:2021lpa,Mohan:2022sxm, Lu:2017meb, Branz:2010pq, Xiao:2017udy}. The magnetic moments and the transition magnetic moments were discussed within the heavy baryon chiral perturbation theory (HBChPT) \cite{Li:2017pxa,Li:2017cfz,Li:2020uok}. 
	
	Significant advancements have been taken with the lattice QCD over the past decade, and some lattice QCD collaborations have simulated the electromagnetic factors of doubly charmed baryons~\cite{Bahtiyar:2022nqw,Can:2013tna,Can:2021ehb,Bahtiyar:2019ykq,Bahtiyar:2018vub}. In addition, the magnetic moments and form factors have been further investigated in the extended on-mass-shell scheme by analyzing lattice QCD results ~\cite{Liu:2018euh,HillerBlin:2018gjw}, and the similar analysis for the transition ones would also be very helpful.   
	
	In this work, we use the unified framework to study both the magnetic form factors $G_{M}(q^2)$ and transition magnetic form factors $G_{M1}(q^2)$ as well as their relevant moments. We study the above quantities from lattice QCD with HBChPT, and finite-volume (FV) effect and lattice spacing correction are specially included and carefully examined. We consider the one-loop contributions and discuss the effects of excited doubly charmed baryons on the magnetic moments of ground ones. Vector mesons are also involved for nonzero $q^2$.
	
	Besides the dimensional regularization, we also employ an alternative finite-range regularization (FRR) to calculate those loop integrals which occur in the HBChPT ~\cite{Donoghue:1998bs,Donoghue:1998krd}. Since the numerical results of these two regularization methods differ very much ~\cite{Donoghue:1998bs,Donoghue:1998krd}, it is important to check whether they are both consistent with the current lattice QCD data. Moreover, the latter regularization is more convenient to study the effect of finite volume. 
	
	%{\color{red}The lattice spacing correction is also considered. It is assumed that corrections for lattice spacings on the order of $a\approx0.1$ fm are small and can be treated as systematic uncertainties~\cite{Ren:2013wxa}. }

	There are three parts in the transition magnetic form factors $G_{M1}(q^2)$, and the $G_3$ relevant terms were ignored when the photon is on shell for the radiative decays of singly heavy baryons and doubly charmed baryons \cite{Li:2017pxa,Wang:2018cre}. As in the $N\to \Delta$ transition process, the $G_3$ terms would also contribute to the $G_{M1}(0)$~\cite{Gellas:1998wx,Gail:2005gz,Faessler:2006ky,Arndt:2003vd, Li:2017vmq}. We will study these terms and investigate their roles in the doubly charmed baryon system.
	
	This paper is organized as follows. In Sec.~\ref{sec:effLag}, we present the contributing effective Lagrangians at $\mathcal{O}(p^2)$.  In Sec.~\ref{3}, the magnetic moments and form factors are studied within HBChPT. The finite-range regularization is used to deal with the loop integrals. The finite-volume and lattice spacing effects are considered. The contributions of vector mesons are introduced to form factors $G_M(q^2)$. In Sec.~\ref{2}, we extrapolate the $G_{M1}(q^2)$ up to $\mathcal{O}(p^3)$ in a similar way with the help of the lattice results. A short summary follows in Sec. \ref{summary}.

	\section{EFFECTIVE LAGRANGIANS}\label{sec:effLag}
	We present the relevant chiral Lagrangians in HBChPT following Refs.~\cite{Scherer:2002tk,Li:2017pxa,Li:2017cfz}. The spin-$\frac{1}{2}$ doubly charmed baryon fields are collected in~\cite{Qiu:2020omj,Liang:2023scp}
	\begin{eqnarray}
		\Psi=\left(\begin{array}{c}
			\Xi^{++}_{cc}\\
			\Xi^{+}_{cc}\\
			\Omega^{+}_{cc}
		\end{array}\right)
		~~~\Rightarrow~~~\left(\begin{array}{c}
			ccu\\
			ccd\\
			ccs\\
		\end{array}\right),
	\end{eqnarray}
	the spin-$\frac{3}{2}$ doubly charmed baryon fields are denoted by the Rarita-Schwinger fields as~\cite{Rarita:1941mf}
	\begin{eqnarray}
		\Psi^{*\rho}=\left(\begin{array}{c}
			\Xi^{*++}_{cc}\\
			\Xi^{*+}_{cc}\\
			\Omega^{*+}_{cc}
		\end{array}\right)^\rho
		~~\Rightarrow~~\left(\begin{array}{c}
			ccu\\
			ccd\\
			ccs\\
		\end{array}\right)^\rho,
	\end{eqnarray}
	and the pseudoscalar meson fields are introduced as 
	\begin{eqnarray}
		\phi=\left(\begin{array}{ccc}
			\pi^0+\frac{1}{\sqrt{3}}\eta&\sqrt{2}\pi^{+}&\sqrt{2}K^{+}\\
			\sqrt{2}\pi^{-}&-\pi^0+\frac{1}{\sqrt{3}}\eta&\sqrt{2}K^{0}\\
			\sqrt{2}K^{-}&\sqrt{2}\bar K^{0}&-\frac{2}{\sqrt{3}}\eta
		\end{array}\right).
	\end{eqnarray}
	We choose the nonlinear realization of the chiral symmetry,
	\begin{eqnarray}
		U=u^2=\exp(i\phi/f_\phi),
	\end{eqnarray}\\
	and $f_\phi$ denote the decay constant of pseudoscalar meson, and $f_\pi=92.4$ and $f_K=113$ MeV are used in this work.
	The chiral axial vector field is defined as~\cite{Scherer:2002tk},
	\begin{eqnarray}\label{eq:umu}
		u_\mu&=&\frac{1}{2}i[u^\dagger(\partial_\mu-ir_\mu)u-u(\partial-il_\mu)u^\dagger],
	\end{eqnarray}\\
	where $r_\mu=l_\mu=-eQA_\mu$, and $Q=$ diag $(2/3,-1/3,-1/3)$ for the pure meson Lagrangians while $Q=$ diag $(2,1,1)$ for baryon fields.
	
	In HBChPT, the charmed baryon fields $\Psi$ can be decomposed into the large component $H$ and the small one $L$,
		\begin{eqnarray}
			H=e^{im_Bv\cdot x}\frac{1+v\hspace{-0.5em}/}{2}\Psi,~~~~~L=e^{im_Bv\cdot x}\frac{1-v\hspace{-0.5em}/}{2}\Psi,
		\end{eqnarray}
		where $v_\mu=(1,\vec{0})$ is the velocity of the baryon. The momentum is decomposed as $p^\mu=m_Bv^\mu+k^\mu$, and thus $k^\mu$ is a small quantity. The HBChPT Lagrangians are obtained from the relativistic Lagrangians after the above separation. In this work, the contribution from small components is beyond the accuracy we consider currently. We use $B$ and $T$ to denote the large components of spin-$\frac{1}{2}$ and spin-$\frac{3}{2}$ doubly charmed baryons. The difference of the mass is denoted $\delta=m_T-m_B$ which appears in the propagators.

	The relativistic  Lagrangian at $\mathcal{O}(p^2)$ contributing to the magnetic moments of doubly charmed baryons at the tree level reads
\begin{eqnarray}\label{La_tree}
			\mathcal{L}^{(2)}_{\gamma BB}=\frac{a_1}{8m_B}\bar \Psi\sigma^{\mu\nu}\hat F^{+}_{\mu\nu}\Psi+\frac{a_2}{8m_B}\bar \Psi\sigma^{\mu\nu}\Psi{\rm Tr}(F^{+}_{\mu\nu}),\nonumber\\\label{p2}
	\end{eqnarray}
	where $a_{1,2}$ are the low-energy constants, and the traceless operator $\hat F^{+}_{\mu\nu}=F^{+}_{\mu\nu}-\frac{1}{3}{\rm Tr}(F^{+}_{\mu\nu})$. The chirally covariant QED field strength tensor $F^{+}_{\mu\nu}=u^{\dagger}F^R_{\mu\nu}u+ uF^L_{\mu\nu}u^{\dagger}$, where $F^R_{\mu\nu}=\partial_\mu r_\nu-\partial_\nu r_\mu-i[r_\mu,r_\nu]$ and $F^L_{\mu\nu}=\partial_\mu l_\nu-\partial_\nu l_\mu-i[l_\mu,l_\nu]$. After the nonrelativistic reduction, the HBChPT Lagrangian is
		\begin{eqnarray}
			\mathcal{L}^{(2)}_{\gamma BB}=a_1\frac{-i}{4m_B}\bar B[S^\mu,S^\nu]\hat F^{+}_{\mu\nu}B+a_2\frac{-i}{4m_B}\bar B[S^\mu,S^\nu]B{\rm Tr}(F^{+}_{\mu\nu}),\nonumber\\
		\end{eqnarray}
		and the $\mathcal{O}(p^2)$ tree-level terms which contribute to the spin-$\frac{1}{2}$ to spin-$\frac{3}{2}$ transition magnetic moment are
		\begin{eqnarray}\label{tree2}
			\mathcal{L}_{\gamma BT}^{(2)} = a_3 \frac{-i}{2m_B} \bar{T}^\mu \hat{F}^+_{\mu\nu} S^\nu B + a_4 \frac{-i}{2m_B} \bar{T}^\mu S^\nu B \mathrm{Tr}(F^+_{\mu\nu}) + \mathrm{H.c.} \, . \nonumber\\
	\end{eqnarray}
	The $\mathcal{O}(p^2) $ pure meson Lagrangian is
	\begin{eqnarray}\label{meson_L}
		\mathcal{L}^{(2)}_{\gamma MM}=\frac{f^2_0}{4}{\rm Tr}[\nabla_\mu U(\nabla^\mu U)^\dagger],
	\end{eqnarray}\\
	with $\nabla_\mu U=\partial_\mu U-ir_\mu U+iUl_\mu.$
	
	The Lagrangian which describes the interaction of vector mesons and photons is given by~\cite{HillerBlin:2018gjw,Borasoy:1995ds}
	\begin{eqnarray}
		\mathcal L_\gamma=-\frac{1}{2\sqrt{2}}\frac{F_V}{m_V}\mathrm{Tr}[V_{\mu\nu}F^{+\mu\nu}]\label{ve1}
	\end{eqnarray}
	where $F_V$ is the decay constant of $V\rightarrow e^+e^-$ and $V_{\mu\nu}=\partial_\mu V_\nu-\partial_\nu V_\mu$ with
	\begin{eqnarray}
		V_\mu=\left(\begin{array}{ccc}
			\frac{1}{\sqrt{2}}\rho^0+\frac{1}{\sqrt{2}}\omega&\rho^{+}&K^{*+}\\
			\rho^{-}&-\frac{1}{\sqrt{2}}\rho^0+\frac{1}{\sqrt{2}}\omega&K^{*0}\\
			K^{*-}&\bar K^{*0}&\phi
		\end{array}\right)_\mu.
	\end{eqnarray}
	
	The $\mathcal{O}(p^1)$ Lagrangian describing the interaction between two baryons and a pseudoscalar meson reads
		\begin{eqnarray}
			\mathcal{L}^{(1)}_{\rm int}&=&\frac{\tilde g_A}{2}\bar\Psi\slashed{u}\gamma_5\Psi+\frac{\tilde g_B}{2}\bar{\Psi}^{*\mu}g_{\mu\nu}\slashed{u}\gamma_5\Psi^{*\nu}\nonumber\\
			&&+\frac{\tilde g_C}{2}[\bar \Psi^{*\mu}u_\mu\Psi+\bar\Psi u_\mu\Psi^{*\mu}].
		\end{eqnarray}
		The corresponding nonrelativistic form is written as
	\begin{eqnarray}\label{bayon_L}
		\mathcal{L}^{(1)}_{\rm int}=\tilde g_A\bar BS^\mu u_\mu B+\tilde g_B\bar T^\rho S^\mu u_\mu T_\rho+\frac{\tilde g_C}{2}[\bar T^\mu u_\mu B+\bar Bu_\mu T^\mu].\nonumber\\
	\end{eqnarray}
	
	To investigate the impact of vector mesons on the magnetic form factors and transition form factors, we refer to Ref.~\cite{HillerBlin:2018gjw} for the following Lagrangians within the HBChPT framework,
    \begin{eqnarray}
		\mathcal L_{VBB}=-\frac{ig_V}{2m_B}\bar{B} [S^\mu, S^\nu] V_{\nu\mu}B,\label{VBB}
	\end{eqnarray}
	\begin{eqnarray}
		\mathcal L_{VBT}=-\frac{i\sqrt{3}d_V}{2m_B}\bar T^{\mu} V_{\mu\nu}S^\nu B.\label{ve2}
	\end{eqnarray}
	
	Utilizing the heavy quark symmetry, the spin-$\frac{1}{2}$ and spin-$\frac{3}{2}$ states can be unified in a superfield~\cite{Falk:1991nq,Meng:2018zbl}
	\begin{eqnarray}
		\psi^{\mu}=T^\mu-\sqrt{\frac{1}{3}}(\gamma^\mu+v^\mu)\gamma_5B.
	\end{eqnarray}
	With the Lagrangian $\kappa\bar{\psi}^\mu\slashed{u}\gamma_5\psi_\mu$ describing the interaction between two baryons and a pseudoscalar meson~\cite{Wang:2018cre}, one can obtain the relations for the pseudoscalar couplings in Eq.~(\ref{bayon_L}):
	\begin{eqnarray}
		\tilde g_B=3\tilde g_A,\quad \tilde g_C=2\sqrt{3}\tilde g_A.
	\end{eqnarray}
	Similarly, the vector couplings satisfy the following relation from the Lagrangian $\displaystyle\frac{i\kappa_V}{m_B}\bar{\psi}^\mu V_{\mu\nu}\psi^\nu$:
	\begin{eqnarray}
		d_V=g_{V}.
	\end{eqnarray}
	\section{magnetic moments and form factors}\label{3}
	\subsection{Magnetic moments}
	In HBChPT, the matrix elements of the electromagnetic vector current for spin-$\frac{1}{2}$ heavy baryon is defined as
	\begin{eqnarray}
		\langle \Psi(p^{\prime})|J_\mu|\Psi(p)\rangle=e\bar u(p^{\prime})\mathcal{O}_{\mu}(p^{\prime},p)u(p).
	\end{eqnarray}
	After obtaining the nonrelativistic limit Lagrangians, the tensor $\mathcal{O}_\mu$ can be expressed as
	\begin{eqnarray}
		\mathcal{O}_\mu(p^\prime,p)=v_\mu G_E(q^2)+\frac{[S_\mu,S_\nu]q^\nu}{m_B}G_M(q^2).
	\end{eqnarray}  
	The magnetic moment and magnetic radii can thus be extracted,
	\begin{eqnarray}\label{mu}
		\mu_B=\frac{e}{2m_B}G_M(0),~~~~\left<r^2_M\right>=\frac{6}{G_M(0)}\frac{dG_M(q^2)}{dq^2}\Bigg|_{q^2=0}.
	\end{eqnarray}
	%----------------------------------------%	
	
	%Ref. \cite{Li:2020uok} shows that next-to-leading order correction with spin-$\frac{3}{2}$ DCB can affect the spin-$\frac{1}{2}$ DCB, 
	
\begin{figure}[htbp]
		\includegraphics[width=80pt]{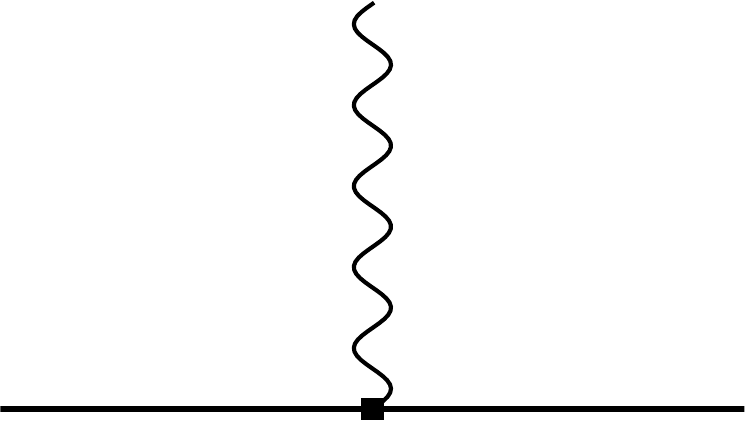}
		\includegraphics[width=80pt]{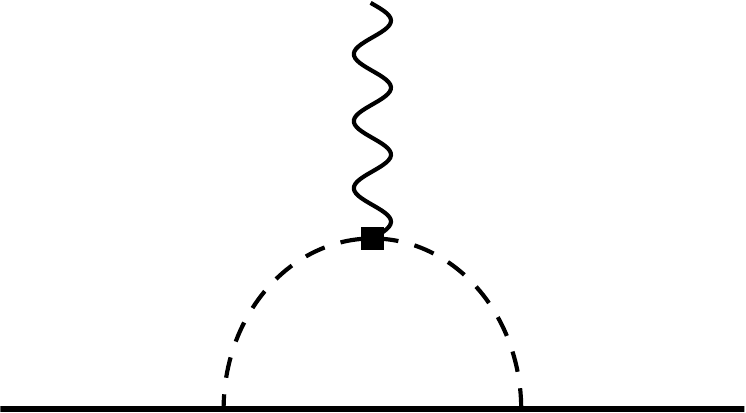}
		\includegraphics[width=80pt]{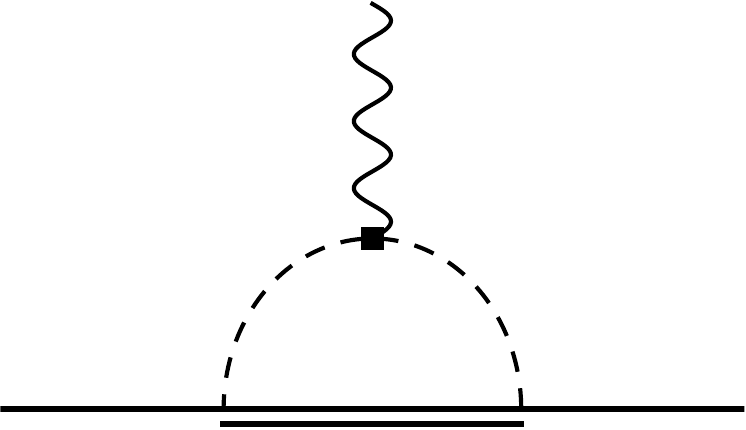}\\
		(a) \qquad\qquad \qquad\qquad
		(b)\qquad\qquad\qquad\qquad(c)
		\caption{Feynman diagrams contributing to the magnetic moment up to $\mathcal{O}(p^3)$ level. The single (double) solid lines refer to the spin-$\frac{1}{2}$ (spin-$\frac{3}{2}$) baryons, and dashed and wavy lines represent the mesons and photons.}\label{tu_mu1}
	\end{figure}

	There are three Feynman diagrams contributing to the magnetic moments up to order $\mathcal O(p^3)$ in Fig.~\ref{tu_mu1}. We also investigated the impact of spin-$\frac{3}{2}$ doubly charmed baryons as the intermediate states through the diagram Fig.~\ref{tu_mu1} (c). The relevant Lagrangians are listed in Sec.~\ref{sec:effLag}. The tree-level diagram comes from the Lagrangian in Eq.~(\ref{La_tree}) which has two couplings $a_1$,~$a_2$. The loop diagrams contain the meson-meson-photon vertices described by Eq.~(\ref{meson_L}) and baryon-baryon-meson vertices with three couplings $\tilde g_A$, $\tilde g_B$ and $\tilde g_C$ as in Eq.~(\ref{bayon_L}). Within the heavy quark limit as detailed in Sec.~\ref{sec:effLag}, the three coupling constants $\tilde g_A$, $\tilde g_B$ and $\tilde g_C$ exhibit well-defined proportionality relations, $\tilde g_B=3\tilde g_A$ and $\tilde g_C=2\sqrt{3}\tilde g_A$. We adopt the value $\tilde{g}_A = -0.4$ obtained with the heavy antiquark diquark symmetry as reported in Refs.~\cite{Sun:2016wzh, HillerBlin:2018gjw, Liu:2018euh}. \footnote{The $\tilde g_A$ in Refs.~\cite{HillerBlin:2018gjw, Liu:2018euh} differs from ours by a factor 2, which arises from a different definition of the $u_\mu$ in Eq. (\ref{eq:umu}).} 
	When the charm quarks are treated as noninteracting spectators in the quark model, the derived coupling relations show agreement with those predicted by the heavy quark limit.
	\begin{table}[htbp]
		\centering
		\caption{The doubly charmed baryons magnetic moments at the tree level.}
		\renewcommand\arraystretch{1.50}
		\setlength{\heavyrulewidth}{1.0pt}
		\begin{tabular}{l|c}%-------------------------------------表格长度，左左中对齐
			\toprule
			\toprule
			$\mu$&\multicolumn{1}{c}{Tree-level expression} \\
			\hline
			$\Xi_{cc}^{++}$& $\frac{e}{m_{\Xi_{cc}^{++}}}(\frac{1}{3}a_1+2a_2)$\\
			$\Xi_{cc}^{+}$& $\frac{e}{m_{\Xi_{cc}^{+}}}(-\frac{1}{6}a_1+2a_2)$\\
			$\Omega_{cc}^{+}$& $\frac{e}{m_{\Omega_{cc}^{+}}}(-\frac{1}{6}a_1+2a_2)$\\
			\bottomrule
			\bottomrule
		\end{tabular}\label{b_mu1}
	\end{table}
	
	The tree-level expression of the magnetic moment is summarized in Table~\ref{b_mu1} and the loop correction is
	\begin{eqnarray}\label{from-loop}
		\mu_B^{(2,{\rm loop})}&=&\sum_{\phi,i}\frac{e\zeta^{(\phi,i)}\tilde g_i^2}{2f^2_\phi}n_3^{\mathrm{III}}|_{q^2\to 0}\\
		&=&\sum_{\phi,i}\frac{e\zeta^{(\phi,i)}\tilde g_i^2}{2f^2_\phi}\frac{-2I_0\omega-J_0+(-2m^2+2\omega^2)L_0}{8},\nonumber
	\end{eqnarray}
	where the $\zeta^{(\phi,i)}$ is listed in Table~\ref{b_mu2}. The loop integrals like $n^{\mathrm{III}}_3$ are defined in the Appendix. When substituting the forms of loop integrals using dimensional regularization into Eq.~(\ref{from-loop}), one can obtain the same results as in Ref.~\cite{Li:2017cfz}.
	
	\begin{table}[htbp]	
		%\raggedright
		\caption{The coefficients of the loop correction in Eq.~(\ref{from-loop}).}
		\renewcommand\arraystretch{1.50}
		\setlength{\heavyrulewidth}{1.0pt}
		\begin{tabular}{l|l|l|l|l}
			\toprule
			\toprule
			&$\zeta^{(\pi,A)}$&$\zeta^{(K,A)}$&$\zeta^{(\pi,C)}$&$\zeta^{(K,C)}$\\
			\hline
			$\Xi_{cc}^{++}$&1&1&$-\frac{1}{3}$&$-\frac{1}{3}$\\
			$\Xi_{cc}^{+}$&$-1$&$0$&$\frac{1}{3}$&0\\
			$\Omega_{cc}^{+}$&0&$-1$&0&$\frac{1}{3}$\\
			\bottomrule
			\bottomrule
		\end{tabular}\label{b_mu2}\\
	\end{table}

	In the discussion below, the momentum-space cutoff is employed to remove the spurious high-energy physics. The contribution of high energies is encoded in the low-energy constants of the local chiral Lagrangian. We use the simple covariant dipole form factor to regulate the contribution integrals~\cite{Donoghue:1998bs,Cloet:2003jm} 
	\begin{eqnarray}\label{dipole_3}
		\left(\frac{-\Lambda^2}{ l^2-\Lambda^2+i\epsilon}\right)^2.
	\end{eqnarray} 
	For example, the integral $J_0(\omega)$ involving one meson and one baryon is regularized as
	\begin{eqnarray}
        &&J_0(\omega)\nonumber\\
        &=&i\int\frac{d^4l}{(2\pi)^4}\frac{1}{(l^2-m^2+i\epsilon)(v\cdot l+\omega+i\epsilon)}\nonumber\\
		&\to&i\int\frac{d^4l}{(2\pi)^4}\frac{1}{(l^2-m^2+i\epsilon)(v\cdot l+\omega+i\epsilon)}\left(\frac{-\Lambda^2}{ l^2-\Lambda^2+i\epsilon}\right)^2\nonumber\\
		&=&\frac{-1}{16\pi^3}\int d^3\vec l\frac{1}{\vec l^2+m^2-\omega\sqrt{\vec l^2+m^2}}\left(\frac{\Lambda^2}{{\vec l}^2+\Lambda^2}\right)^2.\nonumber\\
	\end{eqnarray}

When $\omega\to 0$, the above loop integral is written as 
\begin{equation}
   J_0(\omega\to0)=\frac{-\Lambda^3}{16\pi (\Lambda+m)^2}, 
\end{equation}
which can be expanded in powers of $1/\Lambda$,
\begin{equation}
   J_0(\omega\to0)=-\frac{\Lambda}{16\pi}+\frac{m}{8\pi}+\mathcal{O}(\Lambda^{-1}).\label{J(0)}
\end{equation}
This result is consistent with the one in $d$-dimensional regularization,
\begin{equation}
	J_0(\omega\to0)=4\omega L(\lambda)+\frac{m}{8\pi},\label{nh}
\end{equation} 
  with the well known ultraviolet divergence $L(\lambda)=\frac{\lambda^{(4-d)}}{16\pi^2}\{\frac{1}{d-4}-\frac{1}{2}[\ln (4\pi)+1+\Gamma^\prime(1)]\}$ and the renormalization scale $\lambda$. In this special case, there is no chiral log terms.
  If taking a little complicated case $J_0(\omega\to -m)$ as an example, 
\begin{eqnarray}
			J_0(\omega\to -m)
			&=&\frac{(m^2\pi-2m\Lambda-\pi\Lambda^2)\Lambda}{16\pi^2(\Lambda^2-m^2)}\nonumber\\
   &&+\frac{m\Lambda(2\Lambda^2-m^2)\log{(\frac{\Lambda+\sqrt{\Lambda^2-m^2}}{m})}}{8\pi^2(\sqrt{\Lambda^2-m^2})^3},
   \nonumber\\
\end{eqnarray} 
where the nonanalytic log term is retained in the finite-range regularization. Furthermore, we can check that $J_0(\omega \to -m)$ can give the exact same $m \log m^2$ terms as the one in dimensional regularization when $\Lambda$ is large enough:
\begin{eqnarray}
			J_0(\omega\to -m) &\stackrel{\mathrm{FRR}}{\longrightarrow} &\frac{-\Lambda}{16\pi}-\frac{m}{8\pi^2}\log{(\frac{m^2}{4\Lambda^2/{\rm e}})}+\mathcal{O}(\Lambda^{-1}),\label{n}\nonumber\\
&\stackrel{\mathrm{DR}}{\longrightarrow} &4\omega L(\lambda)-\frac{m}{8\pi^2}\log{(\frac{m^2}{\lambda^2{\rm e}})}.\label{nh1}
\end{eqnarray}

	\subsection{The finite-volume effect}
	
	%Because the chiral physics is dominated by the infrared behavior of loop integrals, and lattice QCD have the dominant effect of discretizing the momenta is to introduce a threshold effect~\cite{Young:2004tb,Young:2002cj,Leinweber:2001ac}. 
	In order to study the lattice QCD simulations, we need to consider the finite-volume effects. To introduce the finite-volume effects, the allowed three-dimensional momentum in the loop integrals is discretized 
	\begin{eqnarray}\label{k}
		\vec l_{\vec n}=\frac{2\pi}{L}\vec n,~~~~\vec n=(n_x,n_y,n_z),
	\end{eqnarray}
	where $n_x,~n_y$ and $n_z$ take natural numbers~\cite{Liang:2022tcj}.

	If the spatial lattice extent $L$ is large enough, we can use approximate spherical symmetry and consider only the degenerate states~\cite{Li:2019qvh}. Then the degeneracy of these states can be calculated by $C_3(n)$, where $n=n^2_x+n^2_y+n^2_z$. Now using these definitions above, we rewrite the continuous integral in momentum space by
	\begin{eqnarray}\label{kn}
		\int d^3\vec l \rightarrow~\left(\frac{2\pi}{L}\right)^3\sum_{n\in \mathbb{N}}C_3(n).
	\end{eqnarray}
	The loop integrals will be actually convergent when $n_{\rm max}>75$.
	
	%The sum of the allowed moment is periodically restricted in Eq. (\ref{k}). For example, the range between $0$ to $\frac{2\pi}{L}$ in $\vec{k}_{\vec{n}}$ doesn't contribute to the loop integrals, but the opposite situation is true for infinite volume. This highlights the indispensable role played by discretization, especially around the physical pion mass.
	
	%In the absence of lattice results for DCB from a variety of lattice volumes, one can but precisely describe the impact of the finite volume on the quark mass dependence on a fixed lattice volume. 
	
	\subsection{The lattice spacing effect}
	
	%The momentum alternative does not take into account the effect of lattice spacing $a$ for extrapolation of lattice calculations. To consider $a$ corrections in the magnetic moments, one should match the lattice theory in the chiral effective theory, as $m_q\ll\Lambda_\chi\ll\frac{1}{a}$~\cite{Bar:2003mh,Beane:2003xv,Tiburzi:2005is,Bar:2004xp,Bunton:2006va}. Then select an power-counting scheme to study order by order, such as
	%\begin{eqnarray}
	%		\epsilon^2 \sim\left\{\begin{array}{l}m_q/\Lambda_\chi\\
		%			a^2\Lambda_\chi^2\\
		%			p^2/\Lambda_\chi^2
		%		\end{array}
	%		\right.,
	%	\end{eqnarray}
%where $\Lambda_\chi$ is the chiral symmetry breaking scale.  We only construct the $\mathcal{O}(a^2)$ terms as the clover term has been taken into account in the lattice results we utilize. It is justifiable to neglect the corrections arising from $a$-dependent correction. 

We also examine the lattice spacing effect in this work. Following Refs. ~\cite{Arndt:2004we,Ren:2013wxa,Tiburzi:2005vy}, we construct the concise Wilson matrix which is proportional to the lattice spacing $a$
\begin{eqnarray}
	\rho_+&=&\frac{1}{2}a(u^\dagger c_q u^\dagger+u c_q u),
\end{eqnarray}
where $c_q$ denotes the Sheikholeslami-Wohlert coefficient matrix which reads $c_q= {\rm diag}(c_{sw}^u,c_{sw}^d,c_{sw}^s)$. The lattice QCD simulation used below gives $c_{sw}^{u/d,s}=1.715$~\cite{Can:2021ehb}. The $\mathcal{O}(a)$ contributions can be canceled by incorporating the clover term into the lattice action~\cite{Bar:2003mh,Ren:2013wxa}.

We can construct the Lagrangian contributing to magnetic moments with the $\rho_+$ operator
\begin{eqnarray}
	\mathcal{L}^{\mathcal{O}(a^2)}_{int}&=&b_1m_Bi\bar B[S^\mu,S^\nu]\hat F^{+}_{\mu\nu}B{\rm Tr}(\rho_+\rho_+)\nonumber\\
	&&+b_2m_Bi\bar B[S^\mu,S^\nu]\{\hat F^{+}_{\mu\nu},\rho_+\}B{\rm Tr}(\rho_+)\nonumber\\
	&&+b_3m_Bi\bar B[S^\mu,S^\nu]\hat F^{+}_{\mu\nu}\rho_+\rho_+B\nonumber\\
	&&+b_4m_Bi\bar B[S^\mu,S^\nu]\rho_+\hat F^{+}_{\mu\nu}\rho_+B\nonumber\\
	&&	+b_5m_Bi\bar B[S^\mu,S^\nu]\rho_+\rho_+\hat F^{+}_{\mu\nu}B.
\end{eqnarray}
The spacing effects are equal for $\mu_{\Xi_{cc}^{+}}$ and $\mu_{\Omega_{cc}^{+}}$
\begin{eqnarray}
	\frac{e}{6}c_{sw}^2a^2m_B(12b_1+24b_2+4b_3+4b_4+4b_5)=\frac{e}{6}c_{sw}^2a^2m_Bb,\label{20}\nonumber\\
\end{eqnarray}
where $b$ is the undetermined combined parameter.

%	We only consider the corrections introduced by tree level in order to avoid completely distorting the physics of the loop integrals by introducing new parameters. The lattice-spacing artifacts can enter through meson mass in loop integrals~\cite{Arndt:2004we,Tiburzi:2005is}, but the corrections to lowest order mesons doesn't break dependencies in Eq. (\ref{mass_kaon}).

\subsection{Numerical results for magnetic moments}
\begin{figure}[htbp]
	\includegraphics[width=121pt]{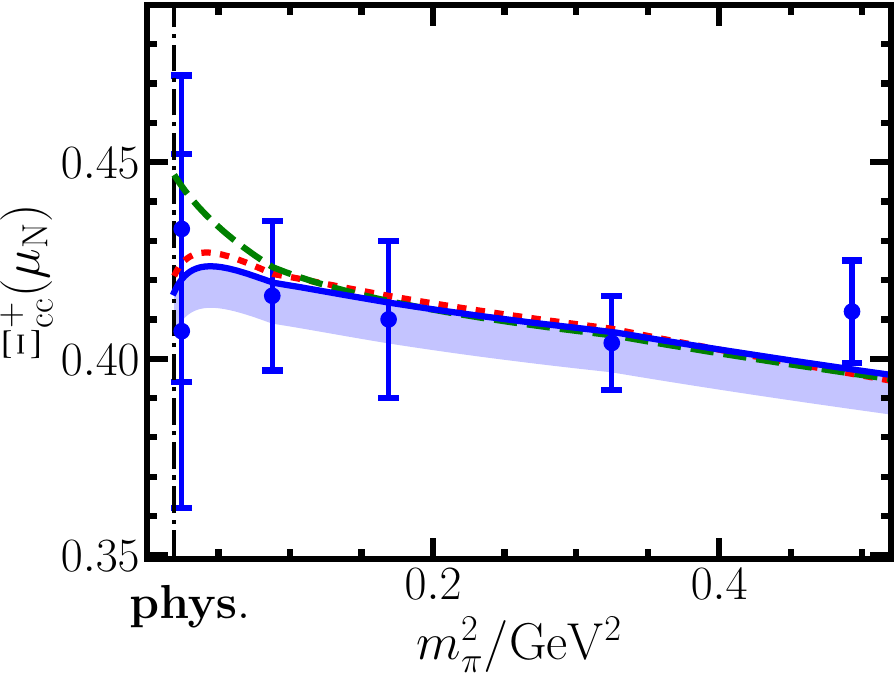}
	\includegraphics[width=121pt]{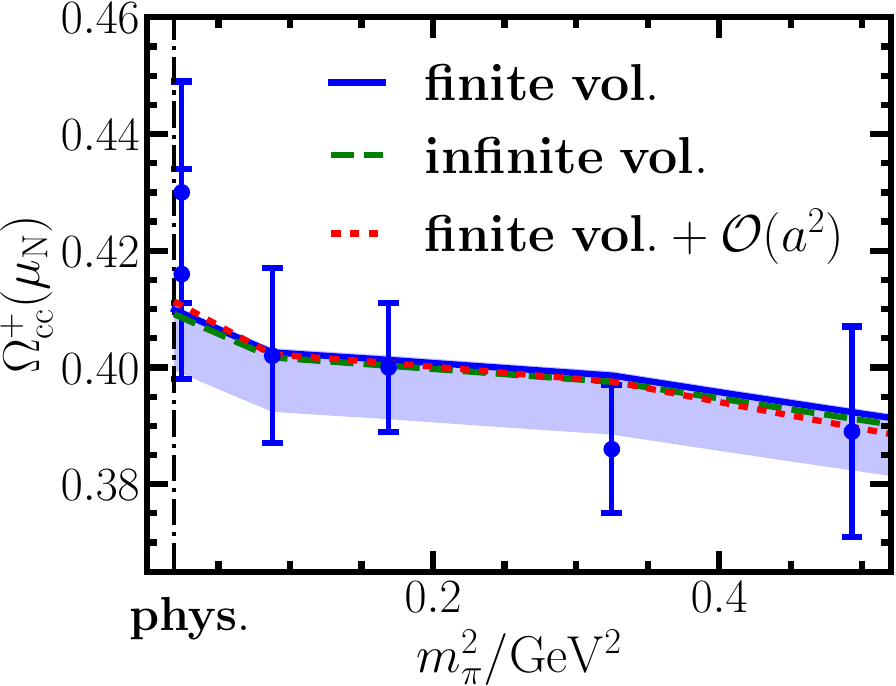}
	\caption{Magnetic moments of $\Xi_{cc}^{+}$ and $\Omega_{cc}^{+}$ as a function of $m_\pi$ in units of $\mu_N$. The lattice results are taken from Refs.~\cite{Can:2021ehb,Bahtiyar:2022nqw} in 2+1 flavor QCD,  shown as the blue points. The green dashed and blue solid curves label our infinite and finite results, and the red dotted curves correspond to the scenario where both discrete momenta and $\mathcal O(a^2)$ are considered. The blue shaded regions illustrate the allowed probable range in our framework when we just fit the lattice results of $m_\pi=0.296,~0.411,~0.570$ GeV.}\label{tu_mu3}
	%The upper two figures show the fitting without considering that the middle state is spin-$\frac{3}{2}$ baryons, while the lower figures show the fitting when this is taken into account. 
\end{figure}

The magnetic moments are given at large pion masses in the lattice QCD simulations as shown in Fig.~\ref{tu_mu3} ~\cite{Can:2021ehb,Bahtiyar:2022nqw}. To relate the kaon and pion masses the following $\chi$PT relation is utilized ~\cite{Liu:2018euh,Cloet:2003jm,Wang:2008vb}
\begin{eqnarray}
	m_K^2=\frac{1}{2}m_\pi^2+(m^2_K-\frac{1}{2}m_\pi^2)_{\rm phys}.\label{mass_kaon}
\end{eqnarray}
We use the lattice QCD masses in Refs.~\cite{Can:2021ehb,Bahtiyar:2022nqw} for other hadrons. 

\begin{table}[htbp]	
	\raggedright
	\caption{The final fitting coupling constants in three cases. }
	%`B\&T' indicates considering the contribution of spin-$\frac{3}{2}$ baryons, while `B' indicates not considering. `Discrete' and `continuous' respectively represent discrete momentum and continuous momentum. And the superscripts `$a^2$' indicate that we consider the effect of $a$.
	\renewcommand\arraystretch{1.50}
	\setlength{\heavyrulewidth}{1.0pt}
	\centering
	\begin{tabular}{l|c|c|c}
		\toprule
		\toprule
		&$b$&$-\frac{1}{3}a_1+4a_2$&$\chi^2$\\
		\hline
		FRR&-&$1.61_{\pm0.02}$&5.02\\
		%B\&T , disc.
		FRR + FV&-&$1.62_{\pm0.02}$&4.30\\
		FRR + FV + $\mathcal{O}(a^2 )$ &$-0.07_{\pm0.23}$&$1.83_{\pm0.68}$&4.20\\
		\bottomrule
		\bottomrule
	\end{tabular}\label{b_mu4}\\
\end{table}

We study these lattice QCD data and examine the finite volume and lattice spacing effects. We choose the cutoff with 0.7 GeV. Three scenarios of fits are given in Fig.~\ref{tu_mu3}, and the corresponding parameters are provided in Table.~\ref{b_mu4}. From the left part of Fig.~\ref{tu_mu3}, one notices that the magnetic moment of $\Xi_{cc}^{+}$ near the physical pion mass seems to deviate a little from the dashed line without the finite-volume effect, which is why the $\chi^2$ is big in the first line in Table.~\ref{b_mu4}. After considering the finite-volume effect, the $\chi^2$ decreases by about 15\%. The finite-volume effect is important for interpreting the lattice QCD data.

The $\chi^2$ is further reduced when the lattice spacing effect is also taken into account from Table.~\ref{b_mu4}, but this effect is small. The parameter for the lattice spacing effect is fitted as 
\begin{equation}
	b=-0.07\pm0.23.
\end{equation}
Its error is larger than the central value, which means the current accuracy in the lattice QCD simulations cannot effectively constrain the lattice spacing effect yet. The dotted and solid lines are also very close in Fig.~\ref{tu_mu3}. Therefore, in the following analysis the lattice spacing effect can be safely neglected.  

%We first study the magnetic moments of $\Xi_{cc}^{+}$ and $\Omega_{cc}^{+}$ in HBChPT by using the dipole form factor to remove the short-distance physics for loop integrals, then we use Eq. (\ref{kn}) and Eq. (\ref{20}) to discuss the finite-volume effects. 

%In Ref.~\cite{Bahtiyar:2022nqw}, the author uses the relativistic fermion action, et Tsukuba action, to obtain the magnetic moments of DCB at $m_\pi\approx156$ MeV in addition to the standard Clover action. This action can reduce the leading cutoff effects which can be removed by tuning the action's parameters~\cite{Bahtiyar:2022nqw,Aoki:2001ra}. 

%	This work only counts the magnetic moments of $\Xi_{cc}^{+}$ and $\Omega_{cc}^{+}$, so we utilize coupling $c=-\frac{1}{3}a_1+4a_2$ instead of $a_1$ and $a_2$. 

%	Our fitting results are depicted in Fig.~\ref{tu_mu3}, the black dashed lines correspond to the continuous-momentum calculation of the loop integrals, the green solid lines correspond to the version of summing the discrete momenta, and the red dotted lines show that the results of both discrete momenta and $a^2$ are considered. We consider or ignore the contribution of spin-$\frac{3}{2}$ baryons to explore the effect of them. 

\begin{figure}[htbp]
	\includegraphics[width=120pt]{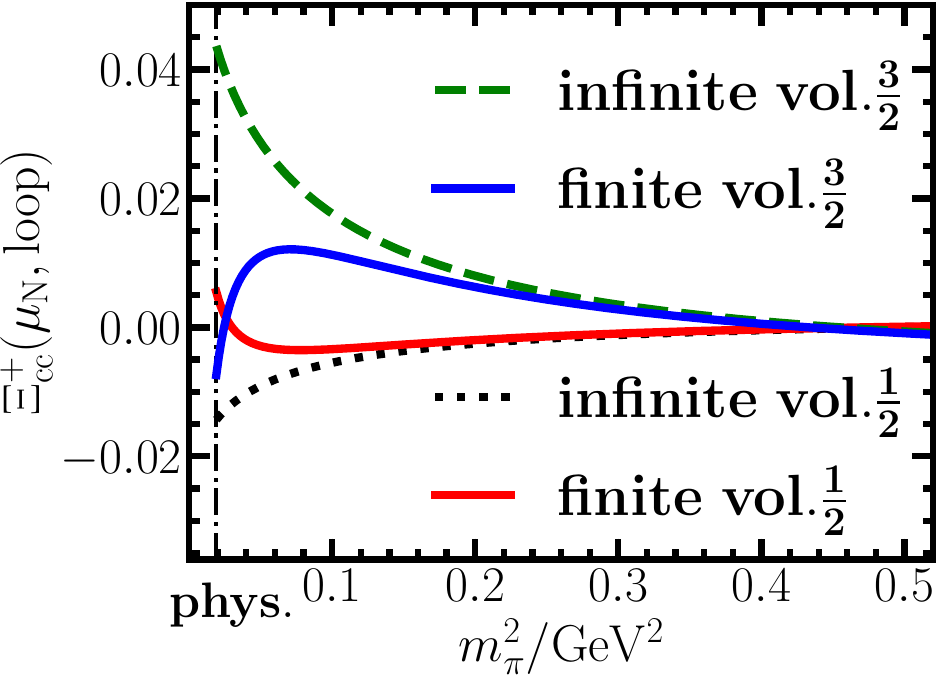}
	\includegraphics[width=123pt]{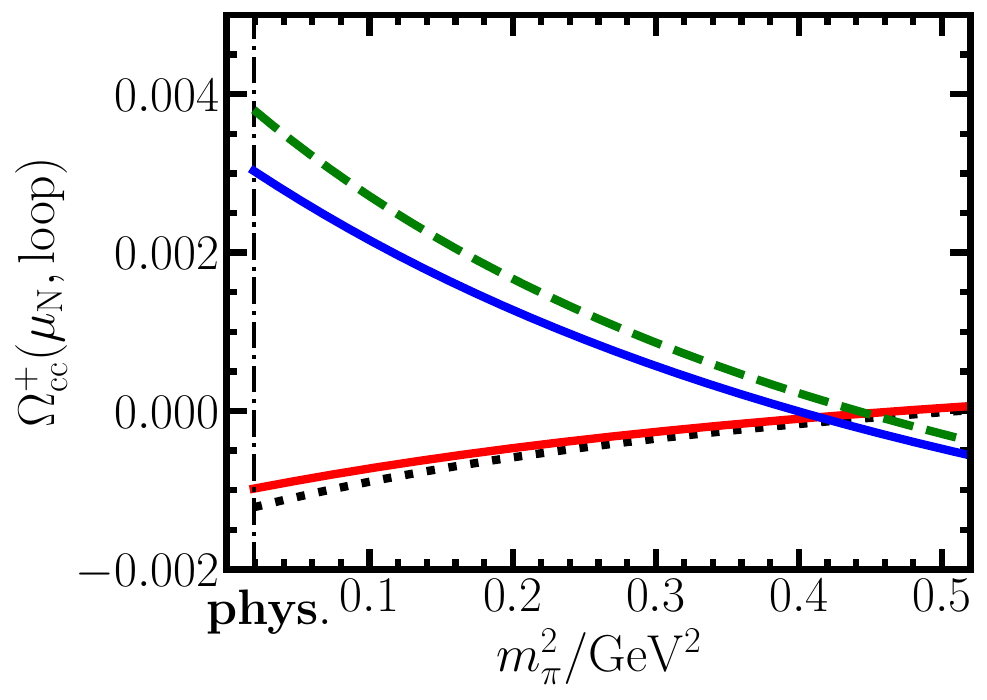}
	\caption{Comparison between the loop contributions to the magnetic moments with the intermediate baryons as the 1/2 and 3/2 doubly charmed baryons.}\label{tu_mu4}
\end{figure}

For a more intuitive representation of the various loop-diagram contributions from different spin baryons, we illustrate the sizes of these contributions in Fig.~\ref{tu_mu4}. The loop-diagram contributions from spin-$\frac{3}{2}$ and spin-$\frac{1}{2}$ doubly charmed baryons exhibit different signs, which is because the coefficients of them have opposite signs as in Table~\ref{b_mu2}. For the same reason, their magnitudes have an approximate ratio of 3:1~\cite{Liu:2012uw}. The graph shows that the main finite-volume correction lies in the chiral limit~\cite{Young:2004tb}.

%The two versions have similar results for DCB in Fig.~\ref{tu_mu3}, but the result of continuous momentum version  fit slightly worse with the lowest two lattice results, including which we think an important result by Tsukube action. The final fitting and extrapolated results are listed in Table~\ref{b_mu4}. From the values of $\chi^2$, the discrete momentum makes $\chi^2$ smaller, regardless of whether the contribution of spin-$\frac{3}{2}$ baryons are taken in to account. Spin-$\frac{3}{2}$ baryons negatively impact the fitting accuracy with respect to the current stage lattice results. The final extrapolated values exhibit greater stability when using the method of summing the discrete momenta. The value is further reduced when the tree-level $a^2$ terms are taken into account, but the effects is small. {\color{red}The tree-level terms for ChPT and lattice spacing are similar that cause the tree-level couplings $c$ and $b$ to be unstable when fitting. Their partial overlap is shown in Table~\ref{b_mu4}.} Improving the accuracy of lattice data would greatly enhance the ability to draw strong conclusions about version determination. 

The dashed and solid lines exhibit an obvious deviation at small pion masses for the magnetic moment of $\Xi_{cc}^+$ in Figs.~\ref{tu_mu3} and \ref{tu_mu4}, particularly when $m_\pi^2<0.1$ GeV$^2$. If we drop the contribution from the loop momentum $\vec l=\vec 0$ in the finite volume, the solid line would approach to the dashed line. The intermediate meson in Fig.~\ref{tu_mu1} is pion for $\mu_{\Xi_{cc}^+}$ while it is kaon for $\mu_{\Omega_{cc}^+}$, which leads to no such phenomenon in the right part because the heavy mass of kaon suppresses the contribution in the finite volume. 

The errors of lattice QCD data near the physical pion mass are relatively large, and the results at the largest pion mass may exceed the applicable range of ChPT extrapolation. Therefore, we refit the lattice data only with $m_\pi=0.296,~0.411,~0.570$ GeV to check whether our results are stable or not. The blue shaded regions in Fig.~\ref{tu_mu3} represent this new fit with $-\frac{1}{3}a_1+4a_2=1.60_{\pm 0.02}$ and $b=0$. As we can see, the results do not change much.

The magnetic moments were studied with the dimensional regularization in Ref.~\cite{Liu:2018euh}, where the four $\mu_{\Xi_{cc}^+}$ at large pion masses on the lattice were fitted without the contributions of the excited doubly charmed baryons in Fig.~\ref{tu_mu1} (c) and that gave $\chi^2=9.77$. We find that $\chi^2$ would become $98.37$ after adding Fig.~\ref{tu_mu1} (c) with the dimensional regularization. However, for the same four data, the $\chi^2$ is only $0.72$ ($1.30$) without (with) Fig.~\ref{tu_mu1} (c) if using the finite-range regularization.

%	The from factor introduce a new variable $\Lambda$. If it's taken so low in energy that it can remove all truly low-energy physics. On the contrary, it can upset the convergence of the  chiral order expansion~\cite{Donoghue:1998bs,Donoghue:1998krd}.  We have chosen $\Lambda=0.8$ GeV $\sim$ 1.0 GeV to prove the extrapolated magnetic moment at physical pion mass is slightly dependent on $\Lambda$. 

In Ref.~\cite{Bahtiyar:2022nqw}, $\mu_{\Xi_{cc}^{++}}/\mu_N=0.089(45)$ is also provided at $m_\pi=0.156$ GeV. With the finite volume effect, we can then get $a_1=-1.26_{\pm 0.17}$ and $a_2=0.30_{\pm 0.01}$ after adding this new datum. 

%With the above couplings constrained by the lattice QCD simulations, the physical magnetic moments in the infinite volume would be predicted in this framework as
%\begin{eqnarray}
%\mu_{\Xi_{cc}^{++}}|_{\rm phy.}&=&0.08_{\pm 0.03}~\mu_N,~\nonumber\\ %\mu_{\Xi_{cc}^+}|_{\rm phy.}&=&0.43_{\pm 0.02}~\mu_N,~\nonumber\\ \mu_{\Omega_{cc}^+}|_{\rm phy.}&=&0.40_{\pm 0.02}~\mu_N.
%\end{eqnarray}

%But $a_1=-2.94$ and $a_2=0.48$ when we fit their results with the help of quark model.
%with the lattice simulations at $m_\pi=0.156$ GeV.

In an effective field theory for low-energy QCD, the magnetic moment can be systematically expanded in powers of $m^2_\pi$, which can be roughly expressed as
\begin{eqnarray}
	\mu_B=c_0+c_1m^2_\pi++I^{({\rm loop})}_{m_\pi}+....
\end{eqnarray}
When regulating the loop integrals using the momentum-space cutoff, the loop integrals generate the obvious cutoff dependence. Taking $J(0)$ in Eq.~(\ref{J(0)}) as an example, the Taylor expansion yields
\begin{eqnarray}
	-\frac{\Lambda}{16\pi}+\frac{m}{8\pi}-\frac{3m^2}{16\pi \Lambda}+\frac{m^3}{4\pi\Lambda^2}+\mathcal{O}(m^4).
\end{eqnarray}
However, the $\Lambda$ variations of the loop diagrams can be absorbed by the redefinitions of the $c_i$ terms from the tree diagrams in principle. 

\begin{figure}[htbp]
	\includegraphics[width=160pt]{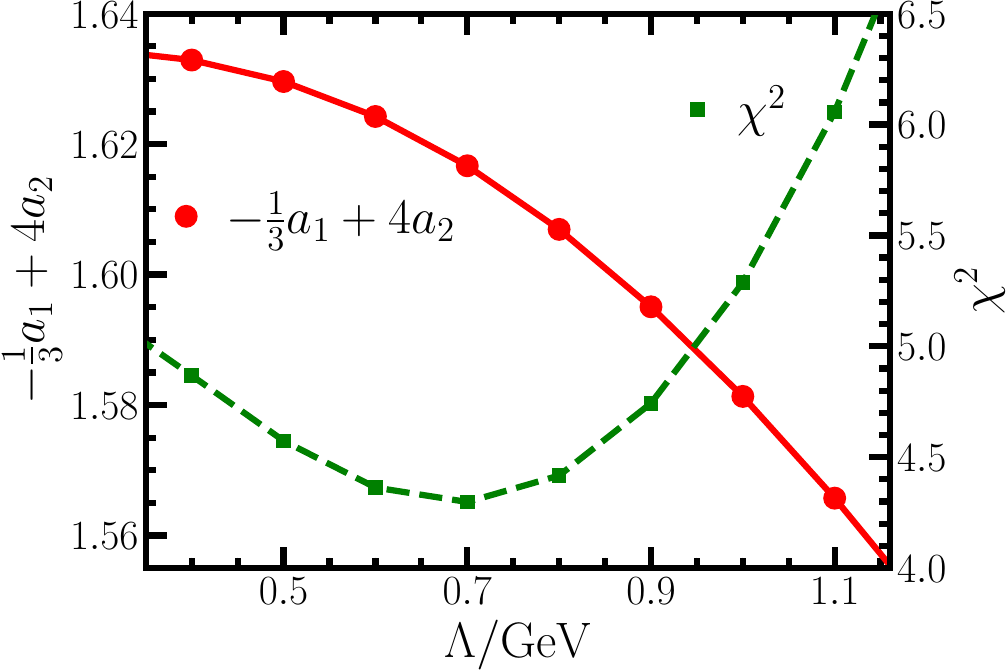}
	\caption{The red dots represent the variation trend of tree-level coupling constants with cutoff, and the green dots are the magnitude of the corresponding $\chi^2$ values. We connect data points with continuous lines for clarity.}\label{tu_x4}
\end{figure}

We refit the above magnetic moments from lattice QCD with different cutoffs. The variation of the coupling $-\frac{1}{3}a_1+4a_2$ is shown in Fig.~\ref{tu_x4}. 
For sufficiently high cutoffs, the fitted coupling displays an approximately linear trend.

To investigate how the fitting quality varies with the cutoff $\Lambda$, we plot the $\chi^2$ values as green dots in Fig.~\ref{tu_x4}. For $\Lambda<1.0$ GeV, the $\chi^2$ is around $5$ and the change is less than $1$, which states the physical observables are insensitive to the choice of the cutoff in this region. However, for $\Lambda>1.0$ GeV, the $\chi^2$ increases significantly, indicating that the convergence of perturbation expansion becomes worse. Since our analysis is limited to one-loop order, incorporating higher-order loops may reduce the cutoff dependence. The optimal fit is achieved at $\Lambda=0.7$ GeV, and thus we adopt this value for our calculations.

Next, we would like to estimate the effect of mass difference $\delta$ on the magnetic moment of spin-$\frac{1}{2}$ charmed baryon. In the above calculation, we fix the mass difference $\delta=0.068$ GeV, which is determined by the lattice QCD data of doubly charmed baryon masses with $m_\pi=0.156$ GeV, as the masses of spin-$\frac{3}{2}$ doubly charmed baryons have not been provided at large pion masses in lattice QCD. Note that the mass difference between $\frac{3}{2}^+$ $\Delta$ and $\frac{1}{2}^+$ nucleon is 0.27-0.33 GeV for pion mass less than 0.6 GeV~\cite{PACS-CS:2008bkb}, and it changes by about 20\%. For the doubly charmed baryon, it contains less light quarks and thus the mass difference may vary less with different pion mass. Accordingly, in the following Fig.~\ref{delta}, we vary the value of $\delta$ as 20\% respective to the central value $\delta=0.068$ GeV and present the magnetic moment $\mu^{\Xi_{cc}^{+}}$ at $m_\pi=0.411$ GeV. The slight change in the blue dashed line indicates that the effect of mass difference is very small.

%\textcolor{red}{Note that the precise masses of spin-$\frac{3}{2}$ doubly charmed baryons have not been provided at large pion masses in LQCD. The mass difference between $\frac{3}{2}^+$ $\Delta$ and $\frac{1}{2}^+$ nucleon is 0.27-0.33 GeV for pion mass less than 0.6 GeV~\cite{PACS-CS:2008bkb}, and it changes about 20\%. For the doubly charmed baryon, it contains less light quarks and thus the mass difference may vary less with different pion mass. We fix the mass difference $\delta$ to its value when $m_\pi=0.156$ GeV. Nevertheless, we should check how much its variation would affect our results. We choose a typical magnetic moment at $m_\pi=0.411$ GeV as shown in Fig.~\ref{delta}. The slight change in the blue dashed line indicates that the effect is very small.} 
	\begin{figure}[htbp]
		\includegraphics[width=220pt]{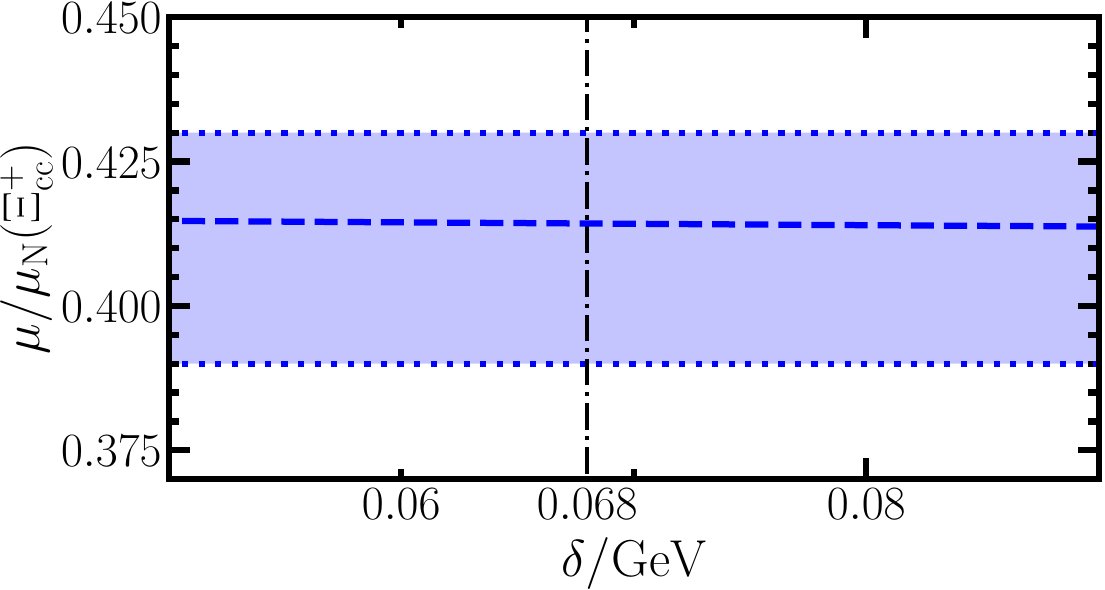}
		\caption{The dependence of the magnetic moment $\mu^{\Xi_{cc}^{+}}$ at $m_\pi=0.411$ GeV on $\delta$, as indicated by the blue dashed line. The blue shaded region represents the error range of the lattice QCD data.} \label{delta}
	\end{figure}

%Let's boldly draw another conclusion. This apparent elimination of cutoff dependence because lattice results are not enough. It will be limited to around $1$ GeV while trying to fit our calculations with the lattice results on the non-physical large pion mass. This conclusion is similar to the results of our previous study on the lattice results of $\Lambda$ baryon energy spectrum~\cite{Liu:2023xvy}. It is essential to study magnetic form factors on non-physical pion masses. If it is indeed impossible to avoid the cutoff in calculating the loop diagram, we must consider what it corresponds to physically and what occurs within the hadrons or during the moment of interaction.

\subsection{Magnetic form factors}
%	The preview fitting results demonstrate the validity of extrapolated lattice results of charmed baryons with respect to the momentum-space cutoff. Deviations from some data points may be caused by various system and fitting errors. The disappointing observation was that the $q^2$ dependence of magnetic moments is poor given the determined $m_\pi$. The problem is common in the electromagnetic form factors of nucleon. 

\begin{figure}[htbp]
	\includegraphics[width=100pt]{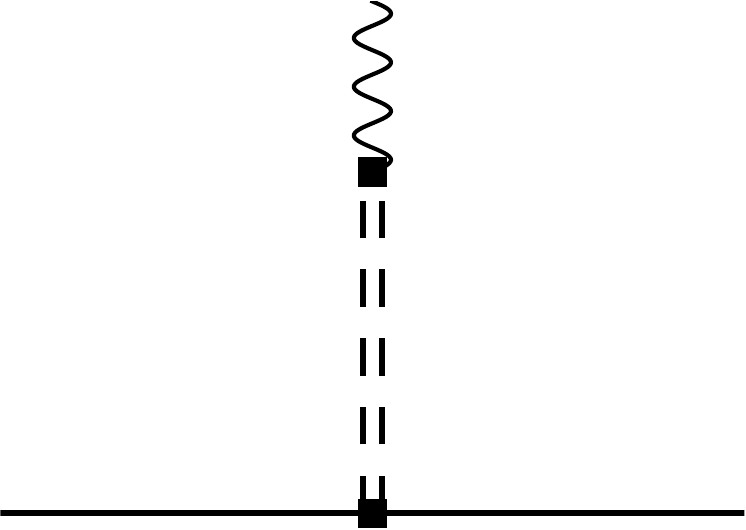}
	\caption{The vector meson dominance diagram for the magnetic form factors of the doubly charmed baryons.}\label{slt}
\end{figure}

To study the magnetic form factors, we need to consider the vector meson contribution as shown in Fig.~\ref{slt}~\cite{HillerBlin:2018gjw}. The Lagrangians involved are listed in Sec.~\ref{sec:effLag}. The results can be expressed as
\begin{equation}
	G_M^B(q^2)=g_V^B\frac{C_{VB} F_V}{m_V}\frac{ q^2}{q^2-m_V^2}.
\end{equation}

Because of the significant breaking of SU(3) symmetry for the vector mesons, two couplings, $g_V^{\Xi_{cc}}$ and $g_V^{\Omega_{cc}}$, are introduced. The coefficients $C_{VB}$ are listed in Table II of Ref.~\cite{HillerBlin:2018gjw}. The above equation shows that the vector mesons do not contribute to $G_M$ as $q^2\rightarrow0$, and that is why we do not mention them for the magnetic moments. The dependence of $G_M$ can also be empirically parametrized as dipole form factor $G_M(Q^2)=G_M(0)/(1+Q^2/0.71~ {\rm GeV^2})^2$~\cite{Kubis:2000zd,Kubis:2000aa,HillerBlin:2017syu,Wang:2007iw,Perdrisat:2006hj,Mergell:1995bf}.

In order to introduce the dependence of vector meson masses on $m_\pi$, we use~\cite{Zhou:2014ila}
\begin{eqnarray}
	m^2_\rho=m^2_\omega&=&m_0^2+2\lambda_m m^2_\pi + \lambda_0(2m_K^2+m_\pi^2),\nonumber\\
	m^2_\phi&=&m_0^2+4\lambda_m m_K^2-2\lambda_m m_\pi^2 + \lambda_0(2m_K^2+m_\pi^2),
\end{eqnarray}
where $m_0\approx0.712$ GeV, $\lambda_m\approx0.489$ and $\lambda_0\approx0.126$. There are also different pion mass dependences for them~\cite{Danilkin:2011fz,Ren:2024frr}, but this discrepancy does not impact our ultimate conclusion.

\begin{figure}[htbp]
	\includegraphics[width=121pt]{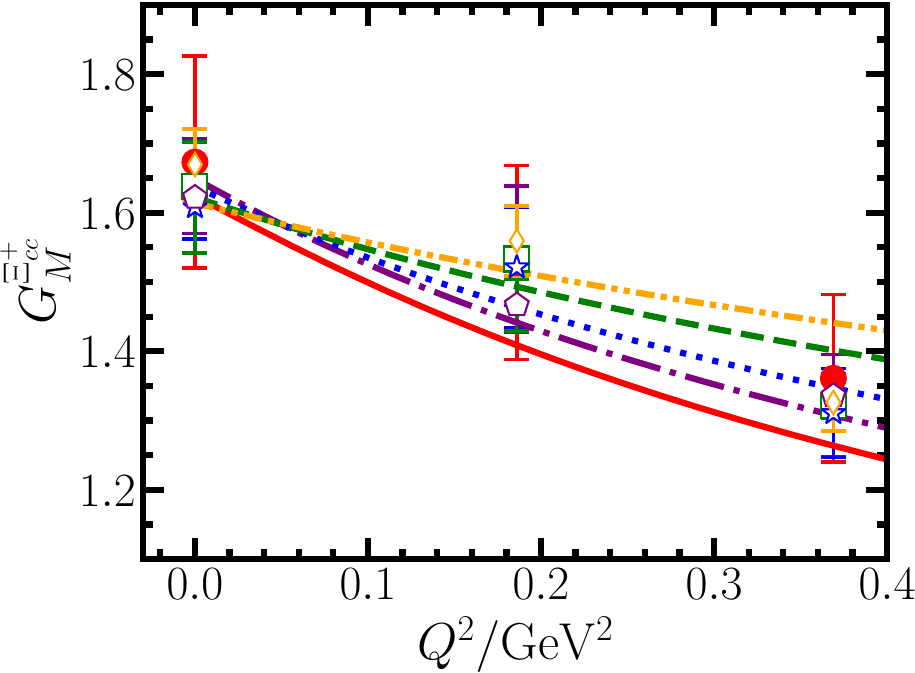}
	\includegraphics[width=121pt]{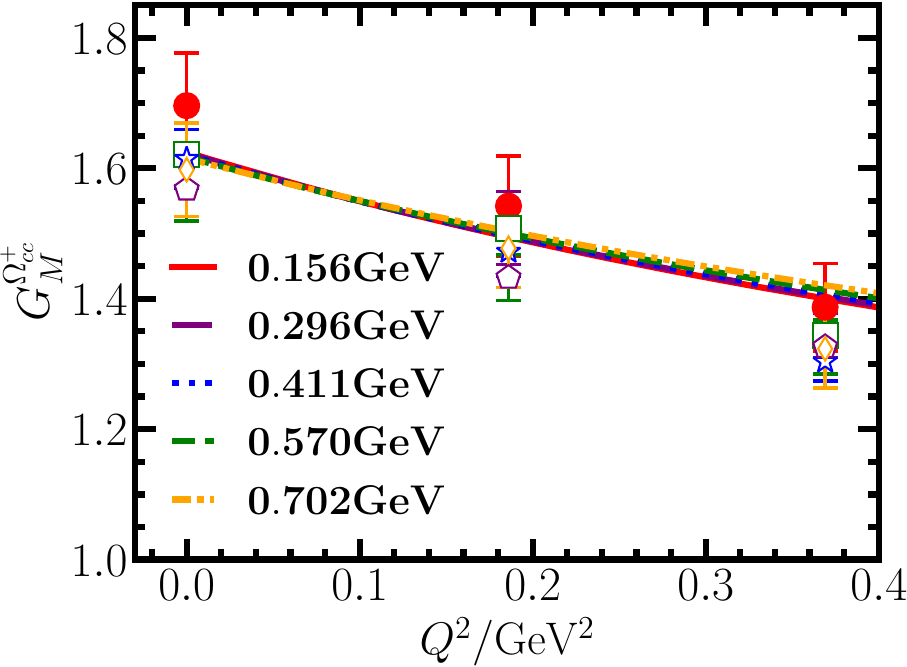}
	\caption{Comparison between the magnetic form factors from the lattice QCD data and our results. The lattice QCD data are taken from Refs.~\cite{Bahtiyar:2022nqw,Can:2021ehb}. The contributions of spin-$\frac{3}{2}$ baryons and the vector meson dominance model in the finite volume are considered. }\label{qq2}
\end{figure}

We obtain the tree-level couplings by fitting the lattice QCD results from Tsukuba actions, where the pion mass is $156$ MeV~\cite{Bahtiyar:2022nqw}, as shown by the red points in Fig.~\ref{qq2}. Both the contributions from vector mesons and discrete-momentum effects are considered. The final results are $g_{V}^{\Xi_{cc}} = 7.95_{\pm 2.04}$ and $g_{V}^{\Omega_{cc}} = 22.84_{\pm 6.09}$.

These parameters are used to calculate the trend as $Q^2$ increases for the other pion masses in Fig.~\ref{qq2}. The different pion masses exhibit almost identical dependence on $Q^2$ for $\Omega_{cc}^+$, as the masses of $c$ and $s$ quarks do not change obviously in the lattice configurations for different pion masses. For $\Xi_{cc}^+$, the left part clearly shows that the slope decreases with the increasing pion mass.

%The lines gradually deviate slightly from the lattice results when the pion mass is large, which meets our expectations.	

%The other uncertainty is that the result is close to the experiments for $Q^2<0.4$ GeV$^2$ when vector mesons are included for the nucleon, but Fig.~\ref{qq2} shows that the range $Q^2<0.8$ GeV$^2$ is still considered valid for DCB. We must remember that there are many unknown systematic errors due to the limited amount of lattice results available. The source of this deviation can be investigated through the dispersion-theoretical analysis in future work.

After obtaining the $g_{V}^B$ and tree-level couplings in Table~\ref{b_mu4}, we can calculate the mean square radius $\left<r^2_M\right>$ in the infinite volume at the physical pion mass. The values are 0.30 fm$^2$ for $\Xi_{cc}^+$ and 0.15 fm$^2$ for $\Omega_{cc}^+$. They are about 1/3 of the $\left<r^2_M\right>_{\rm nucleon}$, similar to those with the extended on-mass-shell scheme in Ref.~\cite{HillerBlin:2018gjw}.

\section{the transition magnetic form factors}\label{2}

\begin{figure}[htbp]
	\includegraphics[width=80pt]{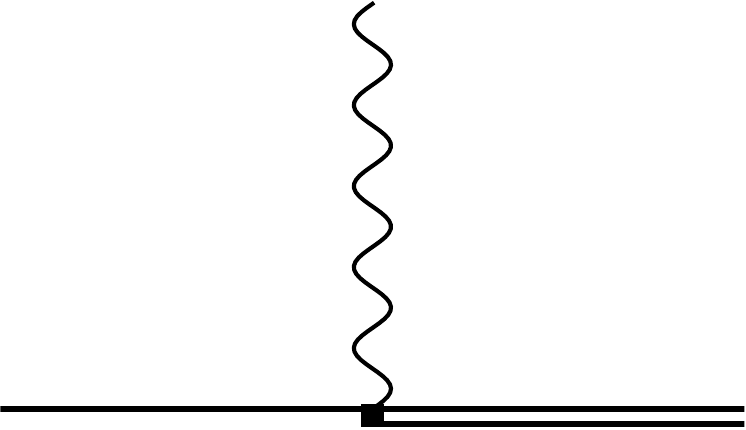}
	\includegraphics[width=80pt]{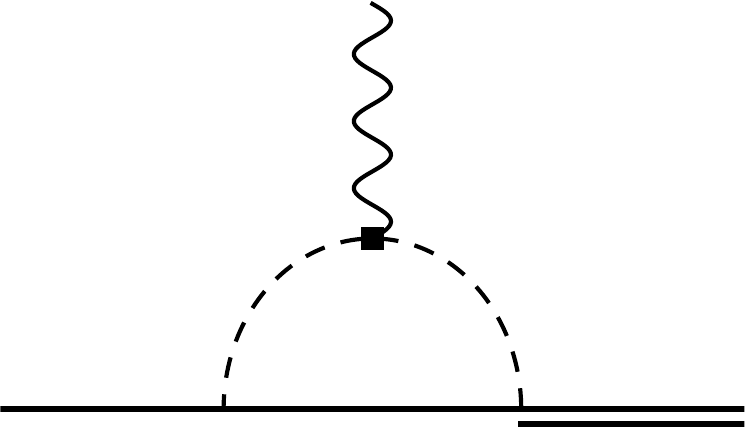}
	\includegraphics[width=80pt]{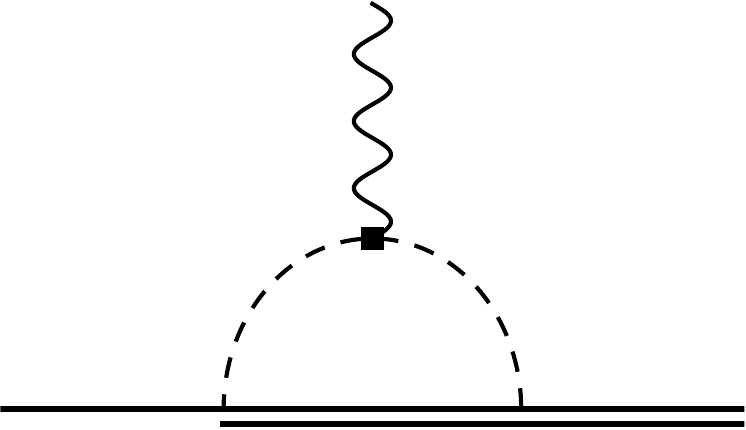}\\
	(a) \qquad\qquad \qquad\qquad
	(b)\qquad\qquad\qquad\qquad(c)
	\caption{Feynman diagrams for the transition magnetic moments.}\label{tu_x1}
\end{figure}

The matrix elements of electromagnetic current between spin-$\frac{1}{2}$ and spin-$\frac{3}{2}$ baryons are defined as~\cite{Jones:1972ky, Bahtiyar:2022nqw}
\begin{eqnarray}
	\langle \Psi^{*\rho}(p^{\prime})|J_\mu|\Psi(p)\rangle=\sqrt{\frac{2}{3}}e\bar u^\rho(p^{\prime})\mathcal{O}_{\rho\mu}(p^{\prime},p)u(p).
\end{eqnarray}
In the framework of HBChPT, the tensor $\mathcal{O}_{\rho\mu}$ can be parametrized in three Lorentz invariant terms
\begin{eqnarray}
	\mathcal{O}_{\rho\mu}(p^{\prime},p)&=&2G_1(q_\rho S_\mu-q\cdot S g_{\rho\mu})\nonumber\\
	&&+G_2\frac{2m_T}{m_B+m_T}(q_\rho v_\mu-q\cdot vg_{\rho\mu})q\cdot S\nonumber\\
	&&+G_3\frac{2}{m_B+m_T}(q_\mu q_\rho-q^2g_{\rho\mu})q\cdot S.\label{duo}
\end{eqnarray}
Here $u^\rho(p)$ is a spin-vector in the Rarita-Schwinger formalism satisfying $v_\rho u^\rho (p)=0 $ and  $\gamma_\rho u^\rho (p)=0$~\cite{Nozawa:1990gt,Hemmert:1997wz,Hemmert:1996xg,Jenkins:1991es,Rarita:1941mf}. The transition magnetic form factor $G_{M1}(q^2)$ is expressed as
\begin{eqnarray}
	G_{M1}(q^2)&=&\frac{m_B}{3(m_T+m_B)}[G_1\frac{(m_T+m_B)(3m_T+m_B)-q^2}{m_T}\nonumber\\
	&&+G_2(m_T^2-m_B^2)+2q^2G_3].
\end{eqnarray}

%\subsection{contributions from pseudoscalar mesons up to $\mathcal{O}(p^2)$}
\subsection{Expressions in the physical world}

There are three Feynman diagrams that contribute to $G_{M1}$ up to $\mathcal{O}(p^3)$, which are depicted in Fig.~\ref{tu_x1}. The tree-level diagram comes from the Lagrangian in Eq.~(\ref{tree2}), while the baryon-meson vertex of the loop diagrams comes from the interaction Lagrangian in Eq.~(\ref{bayon_L}).
The tree diagram only contributes to $G_1$ and can be expressed as
\begin{eqnarray}\label{gm1_tree}
	G_{M1}^{(2,{\rm tree})}=\gamma_B\sqrt{\frac{3}{2}}\frac{(m_T+m_B)(3m_T+m_B)-q^2}{3m_T(m_B+m_T)}.
\end{eqnarray}
The loop contributions are
\begin{eqnarray}
	G_{M1}^{(3,{\rm loop})}=G_{M1}^{(3,b)}+G_{M1}^{(3,c)},
\end{eqnarray}
where 
\begin{eqnarray}
	&&G_{M1}^{(3,b)}=\nonumber\\ \label{gm1_loop}
	&&\sqrt{\frac{3}{2}}\frac{m_B[(m_T+m_B)(3m_T+m_B)-q^2]}{6m_T(m_B+m_T)}\nonumber\\
	&&\times\sum_{i=\pi,K}\left(\alpha_{AC}^{(i)}\frac{\tilde g_A\tilde g_C}{4f^2_i}\right)\left(n_3^{\mathrm{III}}+n_1^{\mathrm{II}}\right)\nonumber\\
	&&+\sqrt{\frac{3}{2}}\frac{(m^2_T-m^2_B)m_B}{6m_T}\sum_{i=\pi,K}\left(\beta_{AC}^{(i)}\frac{\tilde g_A\tilde g_C}{4f^2_i}\right)\left(n_2^{\mathrm{III}}+n_4^{\mathrm{II}}\right)\nonumber\\
	&&+\sqrt{\frac{3}{2}}\frac{m_Bq^2}{3}\sum_{i=\pi,K}\left(\zeta_{AC}^{(i)}\frac{\tilde g_A\tilde g_C}{4f^2_i}\right)\left(2n_1^{\mathrm{III}}+3n_2^{\mathrm{II}}+n_1^{\mathrm{I}}\right),
\end{eqnarray}
and
\begin{eqnarray} \label{GM2c}
	&&G_{M1}^{(3,c)}=\nonumber\\
	&&\sqrt{\frac{3}{2}}\frac{m_B[(m_T+m_B)(3m_T+m_B)-q^2]}{6m_T(m_B+m_T)}\nonumber\\
	&&\times\sum_{i=\pi,K}\left(\alpha_{BC}^{(i)}\frac{\tilde g_B\tilde g_C}{4f^2_i}\right)\left(n_3^{\mathrm{III}}-2n_1^{\mathrm{II}}\right)\nonumber\\
	&&+\sqrt{\frac{3}{2}}\frac{(m^2_T-m^2_B)m_B}{6m_T}\sum_{i=\pi,K}\left(\beta_{BC}^{(i)}\frac{\tilde g_B\tilde g_C}{4f^2_i}\right)\left(n_2^{\mathrm{III}}+n_4^{\mathrm{II}}\right)\nonumber\\	&&+\sqrt{\frac{3}{2}}\frac{m_Bq^2}{3}\sum_{i=\pi,K}\left(\zeta_{BC}^{(i)}\frac{\tilde g_B\tilde g_C}{4f^2_i}\right)\left(2n_1^{\mathrm{III}}+3n_2^{\mathrm{II}}+n_1^{\mathrm{I}}\right).
\end{eqnarray}
We summarize the coefficients $\gamma_B$,          $\alpha_{AC}^{(i)}$ $\alpha_{BC}^{(i)}$, $\beta_{AC}^{(i)}$, $\beta_{BC}^{(i)}$, $\zeta_{AC}^{(i)}$ and $\zeta_{BC}^{(i)}$ in Table~\ref{b_2}. The forms of $n^{\mathrm{I}}_1$, $n^{\mathrm{II}}_1$, $n^{\mathrm{II}}_2$, $n^{\mathrm{II}}_4$, $n^{\mathrm{III}}_1$, $n^{\mathrm{III}}_2$ and $n^{\mathrm{III}}_3$ are also obtained by solving the Lorentzian invariant structures in the Appendix, which have been evaluated in the rest-frame of the spin-$\frac{3}{2}$ baryon~\cite{Gellas:1998wx,Arndt:2003vd}.
\begin{table}[htbp]	
	%\raggedright
	\caption{The coefficients of the tree and loop correction to $G_{M1}$ from Eqs.~(\ref{gm1_tree}) and ~(\ref{gm1_loop}).}
	\renewcommand\arraystretch{1.50}%-------------------------------------------------------设置表格高度
	\setlength{\heavyrulewidth}{1.0pt}
	\begin{tabular}{c|c}
		\toprule
		\toprule
		B&$\gamma_B$ \\
		\hline
		$\Xi_{cc}^{++}$&$-\frac{1}{3}a_3-2a_4$\\
		$\Xi_{cc}^{+}$&$\frac{1}{6}a_3-2a_4$\\
		$\Omega_{cc}^{+}$&$\frac{1}{6}a_3-2a_4$\\
		\bottomrule
		\bottomrule
	\end{tabular}
	\begin{tabular}{c|c|c|c|c|c|c|c|c|c|c|c|c}
		\toprule
		\toprule
		B&$\alpha_{BC}^{(\pi)}$&$\alpha_{AC}^{(\pi)}$&$\alpha_{BC}^{(K)}$&$\alpha_{AC}^{(K)}$&$\beta_{BC}^{(\pi)}$&$\beta_{AC}^{(\pi)}$&$\beta_{BC}^{(K)}$&$\beta_{AC}^{(K)}$&$\zeta^{(\pi)}_{BC}$&$\zeta^{(\pi)}_{AC}$&$\zeta^{(K)}_{BC}$&$\zeta^{(K)}_{AC}$\\
		\hline
		$\Xi_{cc}^{++}$&$-\frac{2}{3}$&$2$&$-\frac{2}{3}$&$2$&$-\frac{2}{3}$&$2$&$-\frac{2}{3}$&$2$&$-\frac{1}{3}$&$1$&$-\frac{1}{3}$&$1$\\
		$\Xi_{cc}^{+}$&$\frac{2}{3}$&$-2$&0&0&$\frac{2}{3}$&$-2$&0&0&$\frac{1}{3}$&$-1$&0&0\\
		$\Omega_{cc}^{+}$&0&0&$\frac{2}{3}$&$-2$&0&0&$\frac{2}{3}$&$-2$&0&0&$\frac{1}{3}$&$-1$\\
		\bottomrule
		\bottomrule
	\end{tabular}\label{b_2}\\
	
\end{table}

The method of deriving loop integrals is universally applicable, and it has been described in detail in the Appendix of Ref.~\cite{Meng:2019ilv}. The results conform to the following relationship,
\begin{eqnarray}
	2(n_1^{\rm II}+n_3^{\rm III})+2(n_2^{\rm III}+n_4^{\rm II})q\cdot v+(2n_1^{\rm III}+3n_2^{\rm II}+n_1^{\rm I})q^2=-2n_3^{\rm III},\nonumber\\
\end{eqnarray}
and this ensures that the numerical matrix elements of electromagnetic current still have the tensor $\mathcal{O}_{\rho\mu}$ structure which can be obtained with the symmetry analysis.

%\subsection{contributions from vector mesons}
\begin{figure}[htbp]
	\includegraphics[width=100pt]{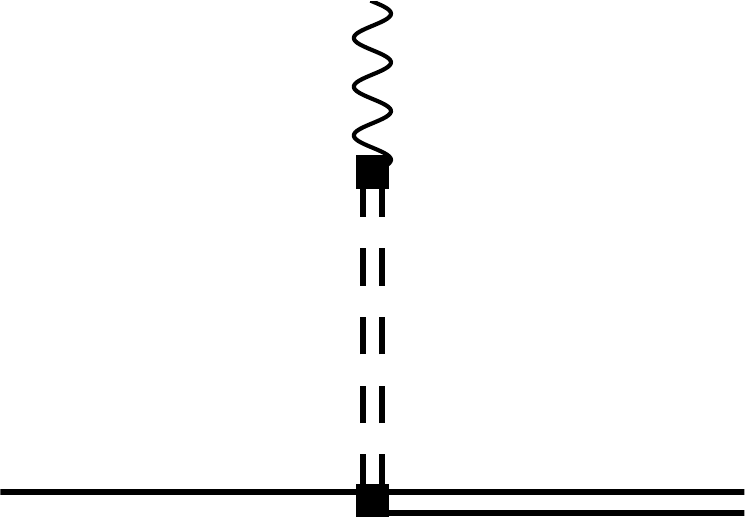}
	\caption{The vector meson dominance diagram for the transition magnetic form factors. %Solid lines are spin-$\frac{1}{2}$ baryons, double solid lines are spin-$\frac{3}{2}$ baryons, dashed lines are mesons, wavy lines are photon and double-dashed lines are vector mesons.
	}\label{tu_x3}
\end{figure}

Similarly, we consider the vector meson contribution as shown in Fig.~\ref{tu_x3} ~\cite{HillerBlin:2018gjw,Aliev:2021hqq,Aliyev:2022rrf}. The specific Lagrangians of the interaction are provided in Eqs.~(\ref{ve1}) and ~(\ref{ve2}). It only contributes through $G_1$ as
\begin{align}
	G_{M1}^{({\rm vec.})}=\xi_V\frac{d^{\Xi_{cc}/\Omega_{cc}}_VF_V}{4\sqrt{2}m_V}\frac{q^2}{q^2-m^2_V}\frac{(m_T+m_B)(3m_T+m_B)-q^2}{m_T(m_B+m_T)},\label{gm1_vec}
\end{align}
where the $\xi_V$ values are listed in Table~\ref{b_3}. 

\begin{table}[htbp]
	\centering
	\caption{The coefficients $\xi_V$ of vector meson contribution in Eq.~(\ref{gm1_vec}).}
	\renewcommand\arraystretch{1.50}
	\setlength{\heavyrulewidth}{1.0pt}
	\begin{tabular}{l|c|c|c}%-------------------------------------表格长度，左左中对齐
		\toprule
		\toprule
		&$\Xi_{cc}^{++}$&$\Xi_{cc}^{+}$&$\Omega_{cc}^{+}$\\
		\hline
		$\rho$&$\frac{\sqrt{2}}{2}$&$-\frac{\sqrt{2}}{2}$&0\\
		$\omega$&$\frac{\sqrt{2}}{6}$&$\frac{\sqrt{2}}{6}$&0\\
		$\phi$&0&0&$-\frac{\sqrt{2}}{3}$\\
		\bottomrule
		\bottomrule
	\end{tabular}\label{b_3}
\end{table}

\subsection{Numerical results}	
%To quantify the agreement with the lattice results and determine the tree-coupling, we use the following $\chi^2$ definition
%\begin{eqnarray}\label{chi2}
%	\chi^2=\sum_{i=1}^{6}\frac{(N_{theo}^i-N_{lattice}^i)^2}{d^2_i}
%\end{eqnarray}
%where $N_{theo}^i$ and $N_{lattice}^i$ are the physical quantities calculated in HBChPT and the lattice QCD and $d_i$ are the error of lattice results.

\begin{figure}[htbp]
		\includegraphics[width=30mm]{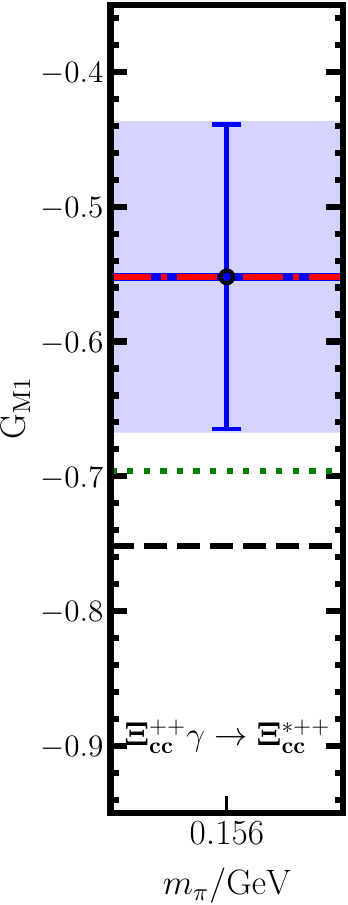}
		\includegraphics[width=24.7mm]{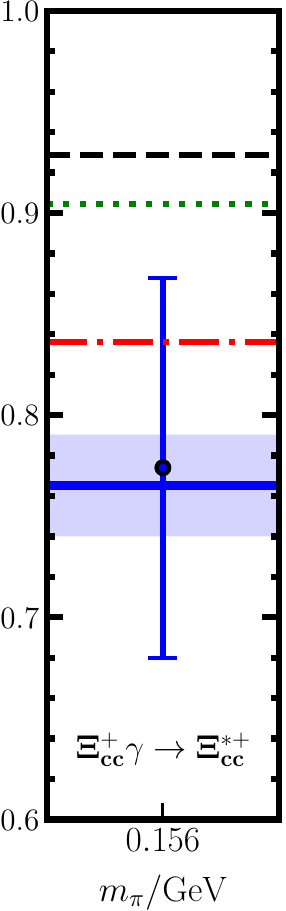}
		\includegraphics[width=24.8mm]{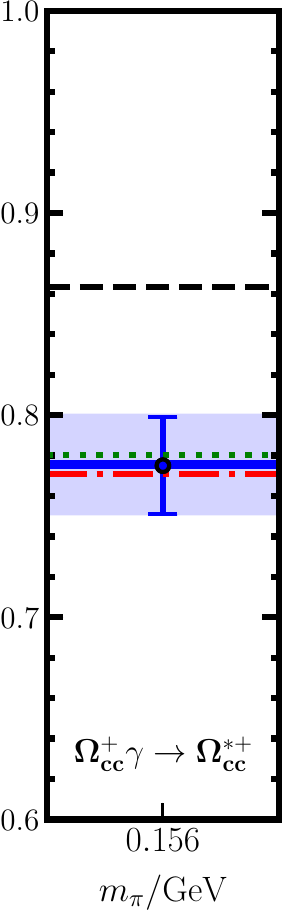}
		\caption{The magnetic multipole transition form factors $G_{M1}$ of doubly charmed baryons at $m_\pi=0.156$ GeV and $Q^2=-q^2\approx0.18$ GeV$^2$ with the spatial lattice extent $L=2.9$ fm. With the corrections of vector mesons, the solid blue lines represent the results in the finite volume while the green dotted lines label our infinite-volume results. The black dashed lines label our finite-volume results for $d_V=0$, that is, the vector meson contributions are turned off. The red dash-dotted lines represent the new fits by adjusting $a_i$ with $d_V=0$. The blue shaded regions illustrate the allowed ranges based on the fit within our framework. The lattice QCD data are taken from Ref.~\cite{Bahtiyar:2018vub}}\label{tu_x2}
	\end{figure}

The lattice QCD simulations provided the multipole form factors of doubly charmed baryons at $m_\pi=0.156$ GeV and the transfer momentum square $Q^2=-q^2\approx0.18$ GeV$^2$ with the spatial lattice extent $L=2.9$ fm in Refs.~\cite{Bahtiyar:2018vub,Bahtiyar:2019ykq}, and we present them in Fig.~\ref{tu_x2}. We explicitly consider the mass splitting, $\delta\approx68$ MeV, between spin-$\frac{1}{2}$ and spin-$\frac{3}{2}$ doubly charmed baryons from the lattice QCD results. 

%	\begin{table}[htbp]
	%	\centering%------------------------------------------------------------------------------------------居中对齐
	%	\caption{The lattice transition form factors for $Q^2=-q^2\approx0.18$ GeV$^2$ and masses of doubly charmed baryons at $m_\pi=0.156$ GeV. There results are taken from the Ref.~\cite{Bahtiyar:2018vub}, which with the spatial lattice extent $L=2.9$ fm.}
	%	\renewcommand\arraystretch{1.50}
	%	\setlength{\heavyrulewidth}{1.0pt}
	%	\begin{tabular}{llll}%-------------------------------------------------------表格长度，左左中对齐
		%		\toprule[1.0pt]
		%		\toprule[1.0pt]
		%		&$\Xi_{cc}^{++}\gamma\rightarrow\Xi_{cc}^{*++}$&$\Xi_{cc}^{+}\gamma\rightarrow\Xi_{cc}^{*+}$ &$\Omega_{cc}^{+}\gamma\rightarrow\Omega_{cc}^{*+}$\\
		%		\hline
		%		$G_{M1}$&-0.552(113)&0.774(94)&0.775(24)\\
		%		\bottomrule
		%		%			\bottomrule[2.0pt]
		%		%\bottomrule
		%		%$m_{\Xi_{cc}}$/GeV&$m_{\Xi_{cc}^*}$/GeV&$m_{\Omega_{cc}}$/GeV&$m_{\Omega_{cc}^*}$/GeV\\	
		%		%\hline
		%		%3.626(30)&3.693(48)&3.719(10)&3.788(11)\\
		%		%\bottomrule[1.0pt]	
		%		\bottomrule[1.0pt]	
		%	\end{tabular}\label{b_1}
	%\end{table}
	
	For the vector meson contributions, we utilize heavy quark symmetry to relate the $TBV$ and $BBV$ couplings as mentioned in the Appendix, and thus $d_{V}^{\Xi_{cc}} =g_{V}^{\Xi_{cc}} = 7.95_{\pm 2.04}$ and $d_{V}^{\Omega_{cc}} = g_{V}^{\Omega_{cc}} = 22.84_{\pm 6.09}$. Our numerical results are presented in Fig.~\ref{tu_x2} with $\Lambda = 0.7$ GeV. The blue lines represent the fit that includes the contributions of vector mesons for the finite-volume version, and the fitted parameters are listed in Table~\ref{b_4}. We also plot the allowed regions of this fit as the blue shades. It clearly shows that the HBChPT with FRR can describe the lattice QCD results well.

	We fix $a_3$ and $a_4$ as in Table~\ref{b_4}, replace the sum of the discrete momenta with the integral of the continuous momenta for the loop diagrams, and obtain the green dotted lines corresponding to the results in the infinite volume. The large difference between the blue solid lines and the green dotted lines indicates that the finite-volume effect is obvious at the current stage.

	To show the effect of the vector mesons, we turn them off by setting $d_V = 0$ and obtain the black dashed lines with $a_i$ in Table~\ref{b_4}. They deviate about $2\sigma$ from the lattice QCD center data, which states that the vector mesons are important for the transition form factors even at small $Q^2=-q^2\approx0.18$ GeV$^2$. Even if we refit the lattice QCD data by adjusting $a_i$ without the vector mesons, the best fit as shown by the red dash-dotted lines exhibits a little tension with the lattice QCD datum in the middle part of Fig.~\ref{tu_x2}.
	
	%Furthermore, the transition couplings $d_V$ of the vector mesons are not adjusted in this section for $G_{M1}$. 

	%	The fitting results support the validity of our interpretation of vector mesons. The solid red lines show a good fit with small dependence on $\Lambda$, highlighting the importance of including vector mesons. Chiral perturbation theory is incomplete without accounting for the contributions of vector mesons. However, due to the limited data available, the advantages of the finite-volume method for extrapolating lattice results are not fully demonstrated, and the calculated results exhibit relatively large errors.
	
	\begin{table}[htbp]	
		\centering
		\caption{The fitted couplings.}
		\renewcommand\arraystretch{1.50}
		\setlength{\heavyrulewidth}{1.0pt}
		\begin{tabular}{c|c}
			\toprule
			\toprule
			$a_3$&$a_4$\\
			\hline
			$1.83_{\pm 0.16}$&$-0.11_{\pm 0.01}$\\
			\bottomrule
			\bottomrule
		\end{tabular}\label{b_4}
	\end{table}

	We separate the $G_{M1}$ into four parts: the tree-level term and the loop terms proportional to $G_1$, $G_2$, and $G_3$. The tree diagram only itself will lead to zero for $G_2$ and $G_3$. Taking the $\Omega_{cc}$ case at $q^2 = -0.18$ GeV$^2$ and $m_\pi = 156$ MeV as an example, 
	\begin{align}
		G^{\Omega^+_{cc}}_{M1}=0.756+[0.019+0.001-0.002]=0.774.
	\end{align}
	It is clear that the tree diagram dominates in our framework, and the loops give about 15\% corrections.

	The $G_3$ related term above is tiny and will be strictly 0 at $q^2 = 0$:
	\begin{align}
		G^{\Omega^+_{cc}}_{M1}(q^2 = 0)=0.843+[0.065-0.062+0]=0.846.
	\end{align} 
	The third lines in Eqs.~(\ref{gm1_loop}) and ~(\ref{GM2c}) correspond to the $G_3$ term and cancel themselves out as $q^2 \rightarrow 0$ due to the constraints from heavy quark symmetry on the couplings $\tilde{g}_A$, $\tilde{g}_B$, and $\tilde{g}_C$ which appear in the loop diagrams. However, with a slight breaking of heavy quark symmetry, for example, $\tilde{g}_A\to \tilde{g}_A$, $\tilde{g}_B\to  0.95 \tilde{g}_B$, and $\tilde{g}_C\to \tilde{g}_C$, one has
	\begin{align}
		q^2G_3(q^2)\,\big|_{q^2\to 0}\neq 0,
	\end{align}
	and 
	\begin{align}
		G^{\Omega^+_{cc}}_{M1}(q^2 = 0)=0.843+[0.061-0.015-0.043]=0.846.
	\end{align}
	If the static $G_3$ terms can be extracted accurately on the lattice, the breaking of the heavy quark symmetry would be understood better.

	%The fitting and extrapolated values are listed in Table~\ref{b_4}. The final extrapolation results in the infinite volume from the continuous-momentum calculation of the loop integrals are
	%\begin{align}
	%	&G_{M1}^{\Xi_{cc}^{++}}(0)=-0.90_{\pm0.09}, \notag\\
	%	&G_{M1}^{\Xi_{cc}^{+}}(0)=1.11_{\pm0.06}, \notag\\
	%	&G_{M1}^{\Omega_{cc}^{+}}(0)=0.89_{\pm0.06}.
	%\end{align}
	%The numerical values can be more credible as more lattice simulation and experimental measurement. 

	\begin{figure}[tb]
		\includegraphics[width=200pt]{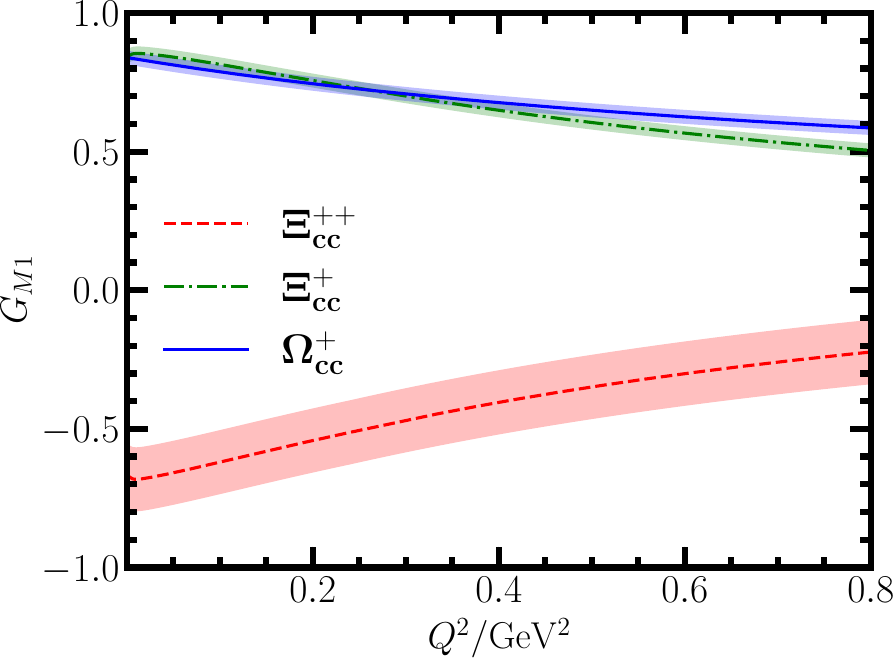}
		\caption{Our predictions for the transition magnetic form factors $G_{M1}$ as $Q^2$ increases with $m_\pi=0.156$ GeV and $L=2.9$ fm. }\label{qq3}
	\end{figure}
	
	Finally, we predict the trend of $G_{M1}$ in Fig.~\ref{qq3} at $m_\pi=0.156$ GeV and $L=2.9$ fm for the range $Q^2=0\sim0.8$ GeV$^2$ with the fitting couplings in Table \ref{b_4}. From the figure, the $G_{M1}^{\Xi_{cc}^{++}}$ exhibits the opposite trend compared to the others. The main reason is that the contributions from the vector mesons have different signs for these transition form factors, which can be easily estimated from Table  \ref{b_3}.
	We hope they can be examined in the future.

	\section{summary}\label{summary}
	
	In this work, we employ HBChPT to study the magnetic form factors and transition magnetic form factors of the doubly charmed baryons up to $\mathcal O(p^3)$. The loop integrals are dealt with the finite-range regularization. We found that finite volume effects are important for interpreting the lattice QCD data, while the lattice spacing effect is not obvious compared to the current accuracy of simulation. We can simultaneously explain three different groups of the lattice QCD data very well as shown in Figs.~\ref{tu_mu3},  \ref{qq2} and \ref{tu_x2}.
	
	We consider both spin-$\frac{1}{2}$ and spin-$\frac{3}{2}$ doubly charmed baryons as the intermediate states in the loop diagrams. For the magnetic moments, the spin-$\frac{3}{2}$ loop contributions are relatively bigger and have different signs from the spin-$\frac{1}{2}$ ones.

	The finite-volume effects improve the consistency of our results with lattice QCD data, particularly in the small pion-mass region. This improvement arises from the contribution of the zero-momentum mode ($\vec{l}=\vec{0}$) in finite volumes. In contrast, the finite-volume corrections for the $\Omega^+_{cc}$ baryon are less pronounced, as the loops involve kaons which are less sensitive to these effects. For the magnetic form factors, the contributions from vector mesons are introduced to examine the $q^2$
	dependence of $G_M$. These corrections become increasingly important as $q^2$ grows.
	
	We have also studied the transition form factors $G_{M1}$ for the processes $\Xi_{cc}^{++}\gamma\rightarrow\Xi_{cc}^{*++}$, $\Xi_{cc}^{+}\gamma\rightarrow\Xi_{cc}^{*+}$ and $\Omega_{cc}^{+}\gamma\rightarrow\Omega_{cc}^{*+}$. The finite-volume effects are also significant for the transition magnetic form factors when fitting to the lattice QCD results. The vector mesons play a crucial role in the transition form factors, even at small momentum transfer, $Q^2=-q^2\approx0.18$ GeV$^2$. We carefully examine the loop contributions associated with the $G_1$, $G_2$, and $G_3$ terms, and the $G_3$ terms becomes bigger if the heavy quark symmetry is not strictly kept. Furthermore, we provide predictions for the transition form factors as a function of $Q^2$.

	We have systematically studied the electromagnetic properties of the doubly charmed baryons within HBChPT, which can also help us understand the nonperturbative strong interactions. We expect our study can be further confirmed in the future lattice QCD calculations or experiments.

	%For magnetic moments, the tree-level coupling constant $-\frac{1}{3}a_1+4a_2=1.58$ is determined from lattice results. The extrapolated infinite-volume results are $\mu_{\Xi_{cc}^{+}}=0.46$, $\mu_N$ and $\mu_{\Omega_{cc}^{+}}=0.40$ $\mu_N$. The agreement with lattice results indicates the necessity of employing discretized momentum. We also observe that the pion loop diagram contributes more significantly than the kaon loop diagram. Furthermore, the inclusion of spin-$\frac{3}{2}$ DCB loops in the fitting process negatively affects the accuracy of the fit with lattice data. When calculating the magnetic form factors, the contributions from vector mesons are introduced to examine the $q^2$ dependence of $G_M$. These corrections become increasingly important as $q^2$ grows. The estimated mean square radius of is about $0.16\sim 0.38$ fm$^2$.

	%We also calculate the transition form factors $G_{M1}$ for the process $\Xi_{cc}^{++}\gamma\rightarrow\Xi_{cc}^{*++}$, $\Xi_{cc}^{+}\gamma\rightarrow\Xi_{cc}^{*+}$ and $\Omega_{cc}^{+}\gamma\rightarrow\Omega_{cc}^{*+}$, and extract $a_3=1.37$, $a_4=-0.12$ by fitting lattice results. The extrapolated values in the infinite volume are $G_{M1}^{\Xi_{cc}^{++}}=-0.98$, $G_{M1}^{\Xi_{cc}^{+}}=1.29$ and $G_{M1}^{\Omega_{cc}^{+}}=0.83$. The contribution of vector mesons are taken into account, and the results indicate that they play an important role in determining the magnetic form factors.

	\section*{ACKNOWLEDGMENTS}
	This project is supported by the National Natural
	Science Foundation of China under Grants No. 12175091, No. 12335001, and No. 12247101, the Fundamental Research Funds for the Central Universities under Grant No. lzujbky-2024-jdzx06, the Natural Science Foundation of Gansu Province under Grants No. 22JR5RA389 and No. 25JRRA799, the ‘111 Center’ under Grant No. B20063, and the innovation project for young science and technology talents of Lanzhou city under Grant No. 2023-QN-107. 
	\appendix
	%\section*{ APPENDIX}
	%\label{app}
	
	\section{ loop integrals }\label{sec:loopInt}
	We collect some common loop functions as follows~\cite{Wang:2018atz,Scherer:2002tk,Meng:2019ilv}
	\begin{eqnarray}
		\Delta&=&i\int\frac{d^4l}{(2\pi)^4}\frac{1}{l^2-m^2+i\epsilon},\\
		I_0(q^2)&=&i\int\frac{d^4l}{(2\pi)^4}\frac{1}{(l^2-m^2+i\epsilon)[(l+q)^2-m^2+i\epsilon]},\nonumber\\\\
		J_0(\omega)&=&i\int\frac{d^4l}{(2\pi)^4}\frac{1}{(l^2-m^2+i\epsilon)(v\cdot l+\omega+i\epsilon)},
	\end{eqnarray}
	\begin{eqnarray}
		i\int\frac{d^4l}{(2\pi)^4}\frac{[1,l_{\alpha},l_{\alpha}l_{\beta},l_{\nu}l_{\alpha}l_{\beta}]}{(l^{2}-m^{2}+i\epsilon)[(l+q)^{2}-m^{2}+i\epsilon](v\cdot
			l+\omega+i\epsilon)} \nonumber\\ =
		[L_{0}(\omega),L_{\alpha},L_{\alpha\beta},L_{\nu\alpha\beta}],~~~~~~~~~~~~~~~~~~~~~~~~~~~~~~~~~~~~~~~~~~~\nonumber\\\label{do}
	\end{eqnarray}
	where
	\begin{eqnarray*}
		L_{\alpha}& = & n^{\rm I}_{1}q_{\alpha}+n^{\rm I}_{2}v_{\alpha},\\
		L_{\alpha\beta} & = &
		n^{\rm II}_{1}g_{\alpha\beta}+n^{\rm II}_{2}q_{\alpha}q_{\beta}+n^{\rm II}_{3}v_{\alpha}v_{\beta}
		+n^{\rm II}_{4}(v_{\alpha}q_{\beta}+q_{\alpha}v_{\beta}),\\		
		L_{\nu\alpha\beta} & = & n^{\rm III}_{1}q_{\nu}q_{\alpha}q_{\beta}+n^{\rm III}_{2}(q_{\nu}q_{\alpha}v_{\beta}
		+q_{\nu}q_{\beta}v_{\alpha}+q_{\alpha}q_{\beta}v_{\nu})\nonumber\\
		&  & 
		+n^{\rm III}_{3}(q_{\nu}g_{\alpha\beta}+q_{\beta}g_{\nu\alpha}+q_{\alpha}g_{\nu\beta})\nonumber\\
		&  & 
		+n^{\rm III}_{4}(q_{\nu}v_{\alpha}v_{\beta}
		+q_{\alpha}v_{\nu}v_{\beta}
		+q_{\beta}v_{\nu}v_{\alpha})\nonumber\\
		&  & +n^{\rm III}_{5}(g_{\nu\beta}v_{\alpha}+g_{\nu\alpha}v_{\beta}
		+g_{\alpha\beta}v_{\nu})+n^{\rm III}_{6}v_{\nu}v_{\alpha}v_{\beta}.
	\end{eqnarray*}

	\vfil

	%参考文献
	
\end{document}